\documentclass[reprint,amsmath,amssymb,aps,showkeys,onecolumn]{revtex4-2}
\usepackage{graphicx}% Include figure files
\usepackage{dcolumn}% Align table columns on decimal point
\usepackage{bm}% bold math
\usepackage{ulem}
\usepackage{amsmath}
\usepackage{amssymb}
\usepackage{mathrsfs}
\usepackage{helvet}
\usepackage{courier}
\usepackage{kbordermatrix}
\usepackage{fancybox,ascmac}
\usepackage[legacycolonsymbols]{mathtools}
\usepackage{color}
\usepackage{mathrsfs}

\begin{document}

%\preprint{APS/123-QED}

\title{Superconfined Antiferromagnons on the Two-Dimensional Penrose Lattice}

\author{Takashi Inoue and Shoji Yamamoto${}^*$}
\affiliation{Department of Physics, Hokkaido University, Sapporo 060-0810, Japan} %\\

\begin{abstract}
   We find novel confined states in the spin-$S$ nearest-neighbor
antiferromagnetic Heisenberg model on the two-dimensional Penrose lattice.
Linear spin waves have massively degenerate eigenstates strictly confined to
tricoordinated sites.
They contrast with the well-known itinerant analogs in the tight-binding model, where
electrons are confined but extended to both tricoordinated and pentacoordinated sites.
It is the site potentials in the spin-wave Hamiltonian, originating from
Coulomb interactions
between electrons, that confine spin waves to minimally coordinated sites only.
Confined states in the tight-binding Hamiltonian consist of six types of building blocks,
whereas those in the spin-wave Hamiltonian consist of only four of them.
Confined spin waves are robust against $1/S$ corrections.
Emergent $O(S^{0})$ interactions further confine---\textit{superconfine}---spin waves into
two separate groups within tricoordinated sites.
%An article usually includes an abstract, a concise summary of the work
%covered at length in the main body of the article. 
%\begin{description}
%\end{description}
\end{abstract}

%\keyword{quasicrystal;
%         Penrose tiling;
%         perpendicular space;
%         Heisenberg antiferromagnet;
%         spin-wave theory;
%         confined state
%        }%Use showkeys class option if keyword
\maketitle

%\tableofcontents

\section{Introduction}
   Quasicrystals have aperiodic but long-range ordered atomic arrangements
\cite{L2477,L596} to exhibit non-crystallographic rotational symmetry and self-similarity.
Since the discovery of such a structure in an Al-Mn alloy \cite{S1951}, numerous transition-metal
alloys have been tried and proved to be quasicrystallined \cite{TL1505,TL1587,T98,T537}.
In particular, magnetic long-range order was reported in rare-earth-based icosahedral
quasicrystals, namely ferromagnetic order in Au-Ga-R (R=Gd, Tb, and Dy) \cite{T19938,T176701}
and antiferromagnetic order in Au-In-Eu \cite{T974}, confirmed by neutron diffraction and
magnetization measurements.
These findings have provided a conclusive answer to the long-standing question of whether
magnetic long-range order can emerge in quasicrystals
\cite{T220201R,I024416,I220403R,Y180409R,S072001}.
Motivated by the observations of magnetic long-range order, as well as quantum criticality
\cite{D1013}, superconductivity \cite{K154}, and related phenomena, theoretical studies based on
ideal models are essential for understanding how aperiodic geometry governs collective
excitations \cite{W043113,S024509,S022002R,T115108,D115413,GL201113,S054202}.

   The $\mathbf{C}_{5\mathrm{v}}$ Penrose tiling \cite{P266} is a two-dimensional realization
of quasiperiodicity.
It not only deserves investigation in itself but also serves as an ideal platform for
exploring quasicrystals especially of icosahedral point group symmetry.
A striking feature of its electronic structure is the macroscopic degeneracy of the
tight-binding Hamiltonian in the vertex model, where electrons hop between adjacent
vertices connected by rhombus edges \cite{C2915,O2184}.
This Hamiltonian hosts a macroscopically degenerate zero-energy eigenlevel
\cite{K2740,A1621,R15827,Z3377,M064213,D064210,K214402}, which originates from a set of
distinctive eigenstates known as \textit{confined states}, whose wavefunctions are
strictly confined to finite domains.
These states are critical in nature---neither localized nor extended in the conventional
sense---as they lack any characteristic length scale.
Arai \textit{et al.} proposed that the confined states are composed of six independent
building blocks \cite{A1621}, which was numerically demonstrated indeed \cite{K214402}.

   Macroscopic zero-energy degeneracy is found in various quasiperiodic
tight-binding models as well, such as the generalized Penrose lattice \cite{O024201},
Ammann-Beenker lattice \cite{K115125,O014204},
Socolar dodecagonal lattice \cite{K360,A064207},
hexagonal golden-mean tilings \cite{K104410,M014413},
quasiperiodic twisted bilayer systems \cite{H165112},
and the recently discovered aperiodic monotile \cite{S086402}.
Confined states do not necessarily occur at zero energy but can emerge at nonzero energies
as well.
In the center model on the Penrose lattice, with an $s$-like orbital placed at the center
of each rhombus, non-interacting electrons exhibit confined states at energy $2t$
\cite{T1420,F2797,T8879}, where $t$ is the hopping integral.
In the $\pi$-flux model on the Ammann-Beenker lattice, confined states appear at energies
$\pm\sqrt{2}t$, $\pm 2t$, and $\pm 2\sqrt{2}t$, as well as at zero energy, depending on
flux configurations \cite{G125104}.

   While such macroscopic degeneracy has been extensively studied in electron hopping models,
analogous phenomena occur in spin systems as well.
The half-filled single-band Hubbard model maps onto a spin-$\frac{1}{2}$ Heisenberg
antiferromagnet in the strong correlation limit \cite{T1289,M9753}.
Quantum Monte Carlo (QMC) simulations reveal that the spatially averaged staggered magnetization
for the Penrose-lattice Heisenberg antiferromagnet is larger than that of the square lattice,
despite both having the same average coordination number of four \cite{J212407,W177205,S11678}.
This suggests that the structural inhomogeneity suppresses quantum fluctuations.
The local magnetizations obtained from linear spin-wave (LSW) theory qualitatively agree with
the QMC results in terms of their coordination-number dependence, and the ground-state energies
from LSW and QMC calculations also agree to within $2\%$ \cite{J212407,S104427}.
Spin dynamics has also been investigated via inelastic photon \cite{I2000118,I053701}
and neutron \cite{Y702} scattering calculations.
Within the harmonic-oscillator approximation for the Heisenberg antiferromagnet on the Penrose
lattice, a prominent peak appears in the density of states at energy $3JS$ \cite{S104427},
where $J$ is the exchange interaction and $S$ is the spin magnitude.
At this energy, numerous degenerate states emerge, characterized by string-like wavefunctions
that form closed loops without a characteristic length scale.
These states manifest as nearly flat-band excitations in the dynamic structure factor,
spanning almost the entire momentum space \cite{Y702}.
A perpendicular-space \cite{D2688} analysis attributes their origin to
antiferromagnons strongly confined to tricoordinated sites.

   We demonstrate that the macroscopically degenerate spin-wave excitations in the Penrose-lattice
Heisenberg antiferromagnet are confined states of antiferromagnons---the magnetic analogs of
the well-known confined states in the tight-binding model.

\section{Model and Method}
\begin{figure}
\centering
\includegraphics[width=0.8\linewidth]{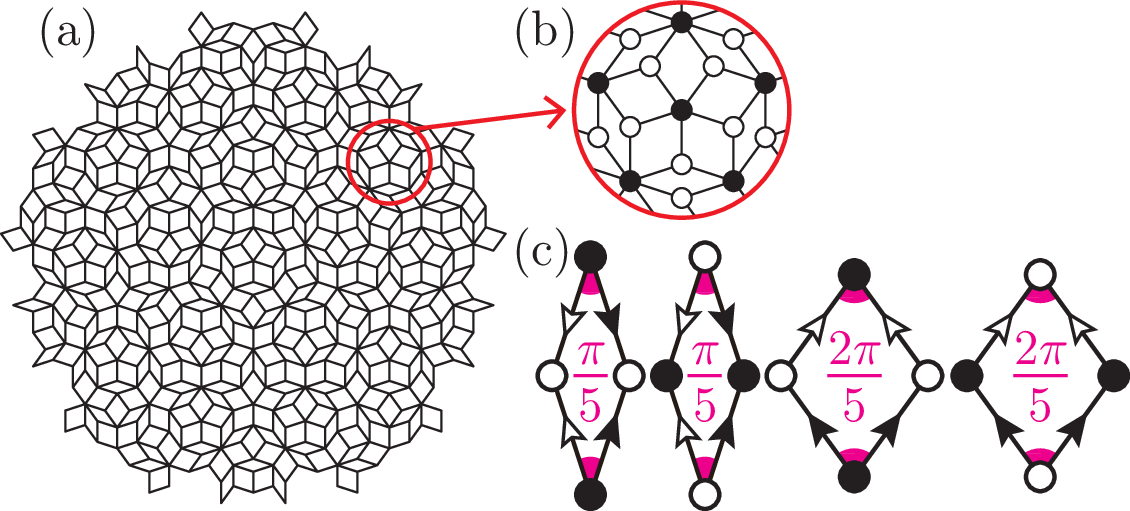}
\caption{%
         (a) A two-dimensional Penrose lattice with size $L=526$ and
         (b) an enlarged view of a specific region.
             In the enlarged view (b), the open and filled circles represent sites belonging to
             the two sublattices.
         (c) Two types of rhombic prototiles constituting the Penrose lattice, with acute
             angles of $\frac{\pi}{5}$ (thin) and $\frac{2\pi}{5}$ (fat).
             Each edge is marked with an arrow to define the matching rules. %add
             For each prototile, there are two possible sublattice assignments at the vertices.
         }
\label{F:Penrosesublattice}
\end{figure}

   We consider the spin-$S$ nearest-neighbor antiferromagnetic
Heisenberg model on the two-dimensional Penrose lattice, which consists of bipartite
sublattices A and B (Figure~\ref{F:Penrosesublattice}).
The Hamiltonian reads
\begin{align}
   \mathcal{H}
  =J\sum_{i\in\mathrm{A}}\sum_{j\in\mathrm{B}}l_{i,j}
   \bm{S}_{\bm{r}_{i}}\cdot\bm{S}_{\bm{r}_{j}}
   \ 
   (J>0).
   \label{E:HeisenbergHam}
\end{align}
We denote the vector spin operators on the A and B sublattices by $\bm{S}_{\bm{r}_{i}}$
($i=1,2,\cdots,L_{\mathrm{A}}$) and $\bm{S}_{\bm{r}_{j}}$ $(j=1,2,\cdots,L_{\mathrm{B}})$,
respectively, where $\bm{r}_{l}$ serves as a generic site label running over both sublattices.
The total number of sites is $L=L_{\mathrm{A}}+L_{\mathrm{B}}$.
The linkage identifier $l_{i,j}$ equals 1 if the vertices $\bm{r}_{i}$ and $\bm{r}_{j}$
are connected, and 0 otherwise.
We express the Hamiltonian~\eqref{E:HeisenbergHam} in terms of the Holstein-Primakoff
bosons \cite{H1098},
\begin{align}
   &
   S_{\bm{r}_{i}}^{+}
  =\left(2S-a_{i}^{\dagger}a_{i}\right)^{\frac{1}{2}}
   a_{i},\ \ 
   S_{\bm{r}_{i}}^{-}
  =a_{i}^{\dagger}
   \left(2S-a_{i}^{\dagger}a_{i}\right)^{\frac{1}{2}},\ \ 
   S_{\bm{r}_{i}}^{z}
  =S-a_{i}^{\dagger}a_{i},
   \allowdisplaybreaks
   \nonumber \\
   &
   S_{\bm{r}_{j}}^{+}
  =b_{j}^{\dagger}
   \left(2S-b_{j}^{\dagger}b_{j}\right)^{\frac{1}{2}},\ 
   S_{\bm{r}_{j}}^{-}
  =\left(2S-b_{j}^{\dagger}b_{j}\right)^{\frac{1}{2}}
   b_{j},\ 
   S_{\bm{r}_{j}}^{z}
  =b_{j}^{\dagger}b_{j}-S,
   \label{E:HPboson}
\end{align}
and expand it in descending powers of $S$ \cite{Y11033},
\begin{align}
    \mathcal{H}
   = \mathcal{H}^{(2)}
    +\mathcal{H}^{(1)}
    +O(S^{0}),
\label{E:Sinvexpansion}
\end{align}
where $\mathcal{H}^{(m)}$, on the order of $S^{m}$, reads
\begin{align}
   &
   \mathcal{H}^{(2)}
 =-JS^2
   \sum_{i\in\mathrm{A}}\sum_{j\in\mathrm{B}}
   l_{i,j},\ 
   \mathcal{H}^{(1)}
  =JS
   \sum_{i\in\mathrm{A}}\sum_{j\in\mathrm{B}}
   l_{i,j}
   \left(
    a_{i}^{\dagger}a_{i}+b_{j}^{\dagger}b_{j}
   +a_{i}b_{j}+a_{i}^{\dagger}b_{j}^{\dagger}
   \right).
\label{E:H2&H1}
\end{align}
We define the up-to-$O(S^1)$ LSW Hamiltonian
\begin{align}
   \mathcal{H}^{[1]}
  \equiv
   \mathcal{H}^{(2)}
  +\mathcal{H}^{(1)}.
   \label{E:LSWHam}
\end{align}

   We introduce the row vectors $\bm{d}^{\dagger}$, $\bm{a}^{\dagger}$, and ${}^{t}\bm{b}$,
of dimension $L$, $L_{\mathrm{A}}$, and $L_{\mathrm{B}}$, respectively, as
\begin{align}
   \bm{d}^{\dagger}
  =\left[a_{1}^{\dagger}, \cdots, a_{L_{\mathrm{A}}}^{\dagger},
         b_{1}, \cdots, b_{L_{\mathrm{B}}}\right]
  \equiv
   \left[\bm{a}^{\dagger}, {}^{t}\bm{b}\right].
\label{E:HProwvector}
\end{align}
Then, we can compactly express the LSW Hamiltonian in matrix form as
\begin{align}
   &
   \mathcal{H}^{[1]}
  =\mathcal{H}^{(2)}
  +\varepsilon^{(1)}
  +JS
   \left[\bm{a}^{\dagger}, {}^{t}\bm{b}\right]
   \left[
   \begin{array}{c|c}
     \mathbf{Z}_{\mathrm{AA}} & \mathbf{C}_{\mathrm{AB}}
    \\ \hline
     \mathbf{C}_{\mathrm{AB}}^{\dagger} & \mathbf{Z}_{\mathrm{BB}}
   \end{array}
   \right]
   \left[
   \begin{array}{c}
     \bm{a}
    \\
     {}^{t}\bm{b}^{\dagger}
   \end{array}
   \right]
   \allowdisplaybreaks
   \nonumber \\
   &
  \equiv
   \mathcal{H}^{(2)}
  +\varepsilon^{(1)}
  +JS\bm{d}^{\dagger}\mathbf{M}\bm{d};\ 
   \varepsilon^{(1)}
  =-JS\sum_{i\in\mathrm{A}}\sum_{j\in\mathrm{B}}l_{i,j},
\label{E:LSWHammatrix}
\end{align}
where $\mathbf{Z}_{\mathrm{AA}}$ and $\mathbf{Z}_{\mathrm{BB}}$ are diagonal matrices
of dimension $L_{\mathrm{A}}\times L_{\mathrm{A}}$ and
$L_{\mathrm{B}}\times L_{\mathrm{B}}$, respectively, and
$\mathbf{C}_{\mathrm{AB}}$ is the biadjacency matrix of dimension
$L_{\mathrm{A}}\times L_{\mathrm{B}}$, as
\begin{align}
   [\mathbf{Z}_{\mathrm{AA}}]_{i,i'}=z_{i}\delta_{i,i'},\ 
   [\mathbf{Z}_{\mathrm{BB}}]_{j,j'}=z_{j}\delta_{j,j'},\ 
   \left[\mathbf{C}_{\mathrm{AB}}\right]_{i,j}=l_{i,j},
\label{E:blockdef}
\end{align}
where $z_{l}$ is the coordination number of the site at $\bm{r}_{l}$.
Denoting the diagonal and off-diagonal blocks of the LSW Hamiltonian matrix $\mathbf{M}$ as
\begin{align}
   \mathbf{Z}
  \equiv
   \left[
   \begin{array}{c|c}
     \mathbf{Z}_{\mathrm{AA}} & \mathbf{O}_{\mathrm{AB}}
    \\ \hline
     \mathbf{O}_{\mathrm{AB}}^{\dagger} & \mathbf{Z}_{\mathrm{BB}}
   \end{array}
   \right],\ 
   \mathbf{T}
  \equiv
   \left[
   \begin{array}{c|c}
     \mathbf{O}_{\mathrm{AA}} & \mathbf{C}_{\mathrm{AB}}
    \\ \hline
     \mathbf{C}_{\mathrm{AB}}^{\dagger} & \mathbf{O}_{\mathrm{BB}}
   \end{array}
   \right],
\label{E:LSWdiagoffdiag}
\end{align}
where $\mathbf{O}_{\mathrm{AA}}$, $\mathbf{O}_{\mathrm{BB}}$, and $\mathbf{O}_{\mathrm{AB}}$
are zero matrices of dimension $L_{\mathrm{A}}\times L_{\mathrm{A}}$,
$L_{\mathrm{B}}\times L_{\mathrm{B}}$, and $L_{\mathrm{A}}\times L_{\mathrm{B}}$, respectively,
we obtain
\begin{align}
   \mathbf{M}
  =\mathbf{Z}
  +\mathbf{T}.
\label{E:LSWmatrix}
\end{align}
We carry out the Bogoliubov transformation \cite{WA450,C327}
\begin{align}
   &
   \bm{d}
  =\left[
   \begin{array}{c|c}
     \mathbf{V} & \mathbf{X}
    \\ \hline
     \mathbf{Y} & \mathbf{W}
   \end{array}
   \right]
   \bm{\alpha};\ 
   [\mathbf{V}]_{i,k_{-}}=v_{i,k_{-}},\ 
   [\mathbf{W}]_{j,k_{+}}=w_{j,k_{+}},\ 
   [\mathbf{X}]_{i,k_{+}}=x_{i,k_{+}},\ 
   [\mathbf{Y}]_{j,k_{-}}=y_{j,k_{-}}
\label{E:Bogoliubov}
\end{align}
with the matrices $\mathbf{V}$, $\mathbf{W}$, $\mathbf{X}$, and $\mathbf{Y}$,
of dimension 
$L_{\mathrm{A}}\times L_{-}$, 
$L_{\mathrm{B}}\times L_{+}$, 
$L_{\mathrm{A}}\times L_{+}$, and
$L_{\mathrm{B}}\times L_{-}$, respectively, to obtain quasiparticle magnons
\begin{align}
   \left[\alpha_{1}^{-\dagger},\cdots,\alpha_{L_{-}}^{-\dagger},
         \alpha_{1}^{+},\cdots,\alpha_{L_{+}}^{+}\right]
  \equiv
   \bm{\alpha}^{\dagger}
\label{E:QPmagnonvector}
\end{align}
and the diagonalized LSW Hamiltonian
\begin{align}
   \mathcal{H}^{[1]}
  =\mathcal{H}^{(2)}
  +\varepsilon^{(1)}
  +\sum_{k_{+}=1}^{L_{+}}\varepsilon_{k_{+}}^{+}
  +\sum_{\sigma=\mp}\sum_{k_{\sigma}=1}^{L_{\sigma}}
   \varepsilon_{k_{\sigma}}^{\sigma}
   \alpha_{k_{\sigma}}^{\sigma\dagger}\alpha_{k_{\sigma}}^{\sigma},
\label{E:LSWdiagHam}
\end{align}
where $\alpha_{k_{\sigma}}^{\sigma\dagger}$ creates a ferromagnetic ($\sigma=-$) or
antiferromagnetic ($\sigma=+$) magnon of energy $\varepsilon_{k_{\sigma}}^{\sigma}$.

   We define the site-resolved density of states 
for the quasiparticle magnon excitation spectrum
\begin{align}
   &
   \rho(\omega)
  =\sum_{l=1}^{L}
   \rho_{l}(\omega)
  =\sum_{z=3}^{7}
   \rho_{z}(\omega)
  =\sum_{z=3}^{7}
   \frac{1}{L}
   \left\{
     \sum_{k_{-}=1}^{L_{-}}
     \frac{ \sum_{i(z_i=z)}\left|v_{i,k_{-}}\right|^{2}
           +\sum_{j(z_j=z)}\left|y_{j,k_{-}}\right|^{2}}
          { \sum_{i\in \mathrm{A}}\left|v_{i,k_{-}}\right|^{2}
           +\sum_{j\in \mathrm{B}}\left|y_{j,k_{-}}\right|^{2}}
     \delta\left(\hbar\omega-\varepsilon_{k_{-}}^{-}\right)
\right.
   \allowdisplaybreaks
   \nonumber \\
   &\qquad\qquad\qquad\qquad\qquad\qquad\quad
\left.
    +\sum_{k_{+}=1}^{L_{+}}
     \frac{ \sum_{i(z_i=z)}\left|x_{i,k_{+}}\right|^{2}
           +\sum_{j(z_j=z)}\left|w_{j,k_{+}}\right|^{2}}
          { \sum_{i\in \mathrm{A}}\left|x_{i,k_{+}}\right|^{2}
           +\sum_{j\in \mathrm{B}}\left|w_{j,k_{+}}\right|^{2}}
     \delta\left(\hbar\omega-\varepsilon_{k_{+}}^{+}\right)
   \right\}.
   \label{E:SWDOS}
\end{align}

\begin{figure}
\centering
\includegraphics[width=0.5\linewidth]{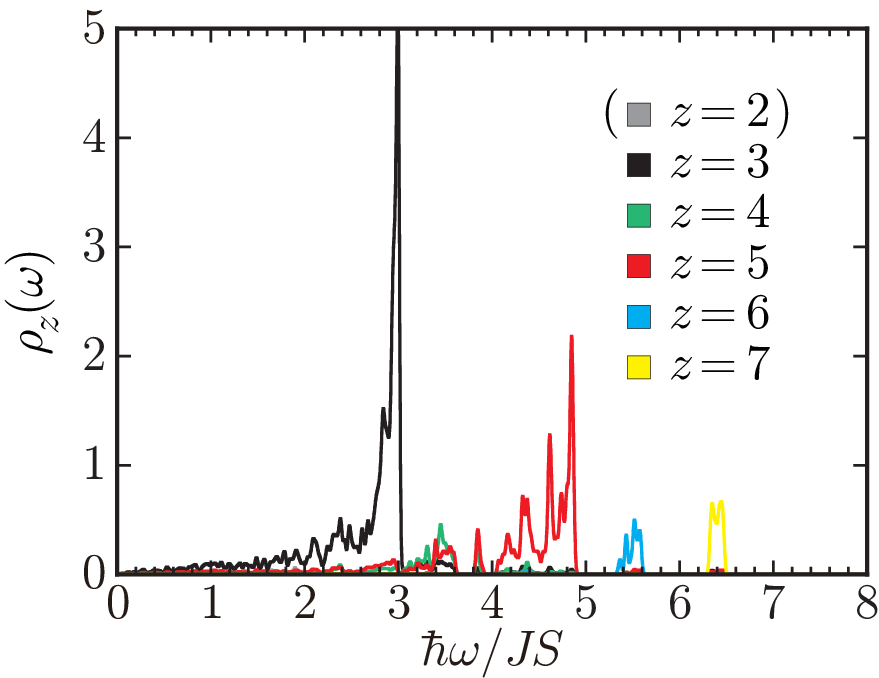}
\caption{%
         The coordination-number-resolved density of states $\rho_{z}(\omega)$ for the LSW
         excitation spectrum on the Penrose lattice with $L=11006$, where $z$ ranges from
         $3$ to $7$.
         Note that $z=2$ vertices are artifacts of finite open-boundary clusters.
         }
\label{F:DOS}
\end{figure}

   Figure~\ref{F:DOS} shows the coordination-number-resolved density of states
$\rho_{z}(\omega)$, where $z$ ranges from $3$ to $7$ (see Appendix~\ref{S:PenroseLattice}).
The spectral weight shifts toward sites with higher coordination numbers as the energy
increases, a feature attributable to the immobility of spin deviations in the Ising limit.
In this limit, the excitation spectrum consists of discrete eigenlevels at $zJS$.
The correlation between coordination number and excitation energy persists even when
the degeneracy is lifted by spin fluctuations.
Notably, the sharp peak at $\hbar\omega=3JS$ indicates macroscopically degenerate confined
states.

\section{Confined Antiferromagnons}
\begin{figure}
\centering
\includegraphics[width=\linewidth]{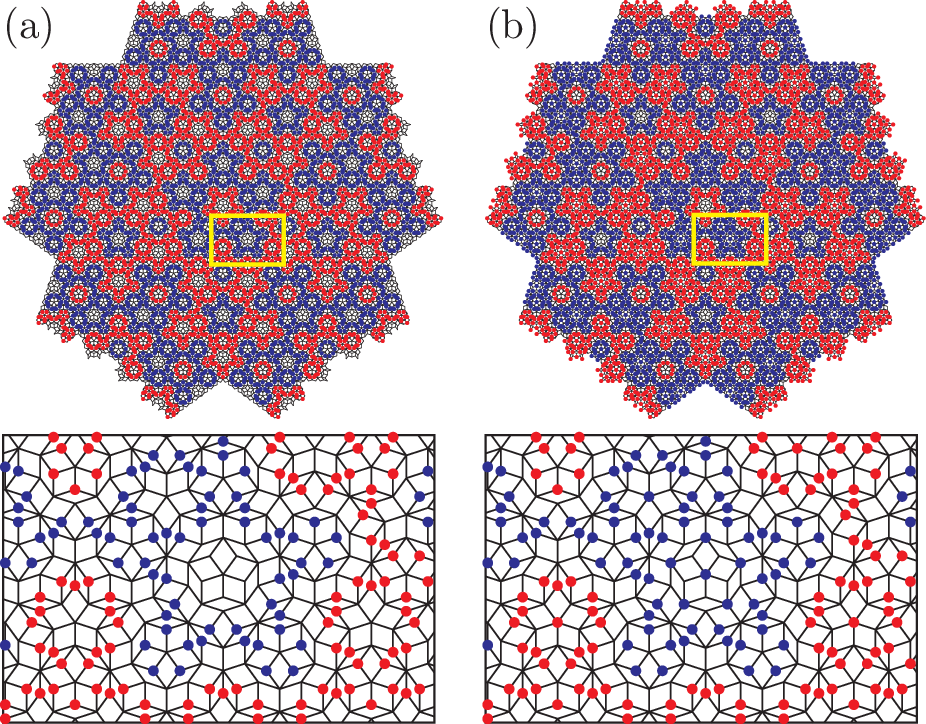}
\caption{%
         The site-resolved densities of states $\rho_{l}(\omega)$ [Equations \eqref{E:SWDOS}
         and \eqref{E:TBDOS}] multiplied by the number of sites $L=11006$ as functions of
         site position $\bm{r}_{l}$.
         The $l$th site is marked with a red ($l\in\mathrm{A}$) or blue ($l\in\mathrm{B}$)
         dot when $L\rho_{l}(\omega)$ is larger than or equal to $10^{-2}$.
         At every unmarked site $l$, $L\rho_{l}(\omega)<10^{-15}$.
         (a) $L\rho_{l}(\omega)$ at $\hbar\omega=3JS$ for the LSW Hamiltonian
             \eqref{E:LSWdiagHam}.
         (b) $L\rho_{l}(\omega)$ at $\hbar\omega=0$ for the tight-binding Hamiltonian
             \eqref{E:TBdiagHam}.
          Rectangles highlight regions enlarged in the bottom panels and
          in Figure~\ref{F:covering}.
         }
\label{F:confinedreal}
\end{figure}

   Figure~\ref{F:confinedreal} demonstrates the spatial confinement of degenerate states
at $\varepsilon_{k}^{\mp}=3JS$ for the LSW Hamiltonian and at $\varepsilon_{k}=0$
for the tight-binding Hamiltonian (see Appendix~\ref{S:TBmodel}) on the Penrose lattice.
Here, $\varepsilon_{k}$ denotes the eigenvalues of the tight-binding Hamiltonian.
In both cases, these confined states are spatially separated by ``forbidden ladders'' 
\cite{A1621,K214402}, which are one-dimensional, unbranched closed loops with zero
wavefunction amplitude.
Within each domain enclosed by these forbidden ladders, the confined states exhibit nonzero
amplitudes exclusively on either the A or B sublattice.
The alternating pattern of confinement domains  across the forbidden ladders matches
perfectly between Figures~\ref{F:confinedreal}a and \ref{F:confinedreal}b.
A key distinction between the two types of confined states lies in their spatial
distributions: in the LSW eigenstates (Figure~\ref{F:confinedreal}a), antiferromagnons
are confined only to tricoordinated sites, whereas in the tight-binding eigenstates
(Figure~\ref{F:confinedreal}b), electrons are confined to tricoordinated and
pentacoordinated sites.

\begin{figure}
\centering
\includegraphics[width=\linewidth]{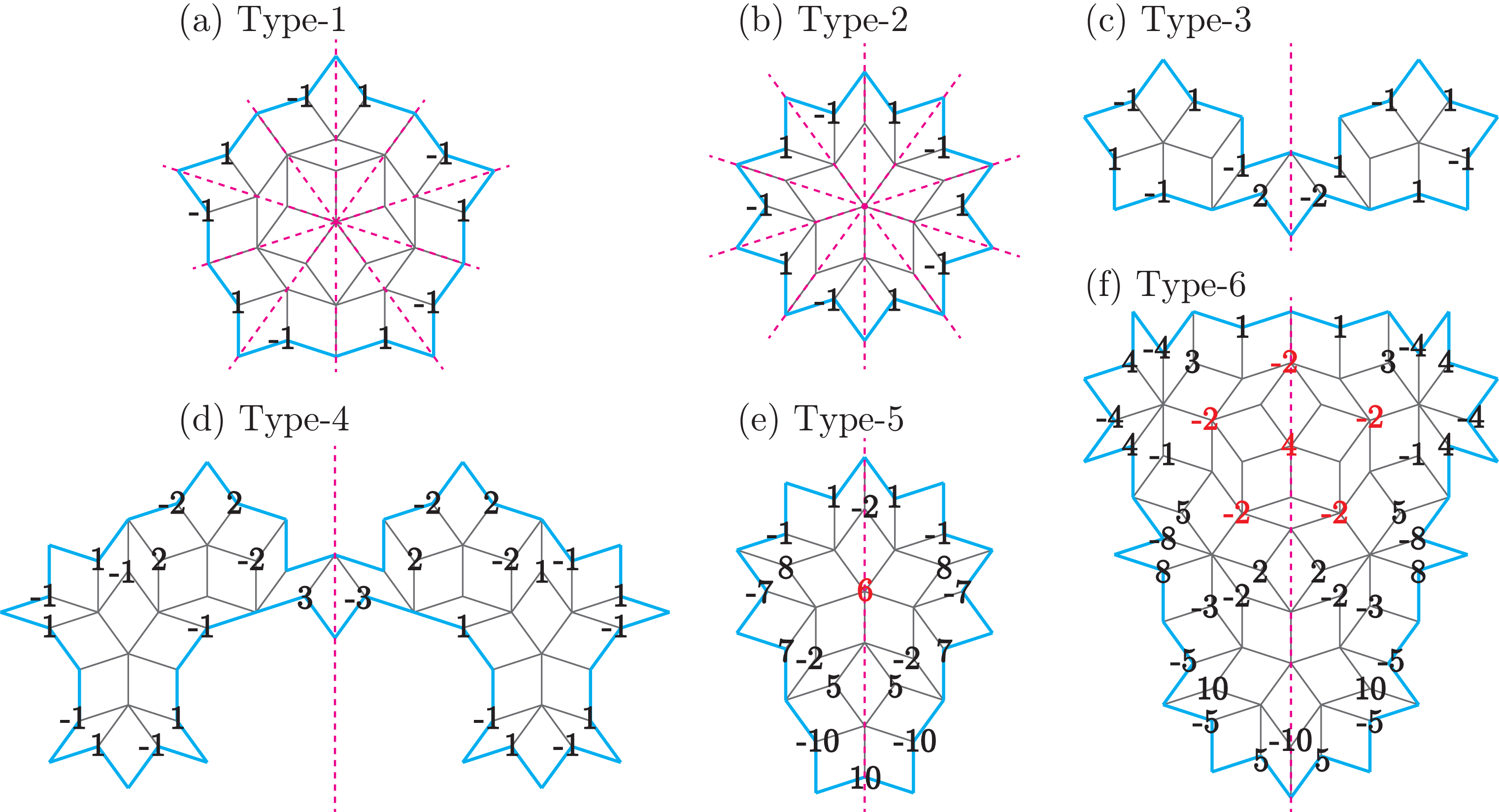}
\caption{%
         Six types of building blocks for the zero-eigenvalue states of the connectivity
         matrix $\mathbf{T}$ on the two-dimensional Penrose lattice.
         Numbers on the vertices indicate the relative amplitude of the unnormalized
         wavefunctions (black: $z=3$ vertices; red: $z=5$ vertices).
         Vertices without numbers have zero amplitude.
         Dotted lines denote mirror-symmetry axes.
         Type-1 to type-4 wavefunctions are mirror-antisymmetric,
         while type-5 and type-6 wavefunctions are mirror-symmetric.
         Type-1 and type-2 building blocks exhibit fivefold rotational symmetry.
         Note that the tiles of type-4 and type-5 include those of type-3 and type-2,
         respectively, and the tiles of type-6 include those of type-1 and type-2.
         }
\label{F:buildingblocks}
\end{figure}

   We derive the constraints on the wavefunction amplitudes of the confined states.
The matrix form of the LSW Hamiltonian~\eqref{E:LSWHammatrix} reduces the problem to
an eigenvalue problem for the matrix $\mathbf{M}=\mathbf{Z}+\mathbf{T}$.
We first examine the off-diagonal connectivity matrix $\mathbf{T}$, which corresponds to
the tight-binding Hamiltonian with the hopping integral set to unity.
Using the argument from the tight-binding model (see Appendix~\ref{S:TBmodel}), we obtain
the local constraints at each site $\bm{r}_{l}$ for the LSW wavefunction~\eqref{E:Bogoliubov}
to be an eigenvector of $\mathbf{T}$ with zero eigenvalue, which read
\begin{align}
   \begin{cases}
     \sum\limits_{j\in\mathrm{B}}l_{i,j}y_{j,k_{-}}=0 &
     (l=i\in\mathrm{A})
    \\
     \sum\limits_{i\in\mathrm{A}}l_{i,j}v_{i,k_{-}}=0 &
     (l=j\in\mathrm{B})
   \end{cases}
\label{E:E3JSconst1}
\end{align}
for the $\sigma=-$ mode and
\begin{align}
   \begin{cases}
     \sum\limits_{j\in\mathrm{B}}l_{i,j}w_{j,k_{+}}=0 &
     (l=i\in\mathrm{A})
    \\
     \sum\limits_{i\in\mathrm{A}}l_{i,j}x_{i,k_{+}}=0 &
     (l=j\in\mathrm{B})
   \end{cases}
\label{E:E3JSconst2}
\end{align}
for the $\sigma=+$ mode.
As established in Refs.~\cite{A1621,K214402}, for the connectivity matrix $\mathbf{T}$ on
the infinite Penrose lattice, the only local solutions satisfying Equations~\eqref{E:E3JSconst1}
and \eqref{E:E3JSconst2} are the six types of building blocks shown in
Figure~\ref{F:buildingblocks}.
We define the $L$-dimensional vector
\begin{align}
   \bm{\psi}_{n,m}
  =\left[
   \begin{array}{c}
     \psi_{n,m}(\bm{r}_{1})
    \\
     \vdots
    \\
     \psi_{n,m}(\bm{r}_{L})
   \end{array}
   \right],\ 
   [\bm{\psi}_{n,m}]_{l}
  =\psi_{n,m}(\bm{r}_{l}),
\label{E:Ldimvector}
\end{align}
where $n=1,\cdots,6$ labels the type of building block, $m=1,\cdots,N_{n}$ labels each
block of type-$n$, and $N_{n}$ is the total number of such blocks.
Each $\bm{\psi}_{n,m}$ is an unnormalized eigenvector of $\mathbf{T}$ satisfying
$\mathbf{T}\bm{\psi}_{n,m}=0$.
The wavefunction amplitudes $\psi_{n,m}(\bm{r}_{l})$ at each site take the values
specified in Figure~\ref{F:buildingblocks} for the corresponding tiles of the Penrose lattice
and are zero elsewhere.
The number of nonzero components for each type of building block is summarized in
Table~\ref{T:numberofsites}.

\begin{table}
\caption{%
         Building-block properties.
         }
\centering
\begin{ruledtabular}
\begin{tabular}{l cccccc}
Type & 1 & 2 & 3 & 4 & 5 & 6 
\\
\hline
Number of constituent sites &
$36$ &
$31$ &
$28$ &
$66$ &
$36$ &
$81$
\\
Number of finite-amplitude sites & 
$10$ &
$10$ &
$12$ &
$28$ &
$19$ &
$41$
\\
\end{tabular}
\end{ruledtabular}
\label{T:numberofsites}
\end{table}

   Next, we examine the effect of the diagonal local potential matrix $\mathbf{Z}$ on the
vectors $\bm{\psi}_{n,m}$.
Applying the matrix $\mathbf{M}$ to the vector $\bm{\psi}_{n,m}$ yields
\begin{align}
   \mathbf{M}\bm{\psi}_{n,m}
  =\mathbf{Z}\bm{\psi}_{n,m}+\mathbf{T}\bm{\psi}_{n,m}
  =\mathbf{Z}\bm{\psi}_{n,m}.
\label{E:eigenEqM}
\end{align}
Since $\mathbf{Z}$ is a diagonal matrix whose elements are the coordination numbers $z_{l}$
at each site $\bm{r}_{l}$, we obtain 
\begin{align}
   [\mathbf{Z}\bm{\psi}_{n,m}]_{l}
  =z_{l}\psi_{n,m}(\bm{r}_{l}).
\label{E:eigenEqZ}
\end{align}
When the coordination number $z_{l}$ is uniform ($z_{l}=z$) across all sites with nonzero
amplitudes, $\bm{\psi}_{n,m}$ is a simultaneous eigenvector of $\mathbf{Z}$ and $\mathbf{M}$,
such that $\mathbf{Z}\bm{\psi}_{n,m}=z\bm{\psi}_{n,m}$ and
$\mathbf{M}\bm{\psi}_{n,m}=z\bm{\psi}_{n,m}$.
Therefore, the type-1 to type-4 building blocks, in which all sites with nonzero
wavefunction amplitude have coordination number $z=3$, are eigenvectors of $\mathbf{M}$
with eigenvalue $3$: %, satisfying
\begin{align}
   \mathbf{M}\bm{\psi}_{n,m}=3\bm{\psi}_{n,m}\ 
   (n=1, 2, 3, 4).
\label{E:eigenEqME3JS}
\end{align}
In contrast, the type-5 and type-6 building blocks, which contain both $z=3$ and $z=5$
sites with nonzero wavefunction amplitudes, are eigenvectors of $\mathbf{T}$ but not of
$\mathbf{M}$.
Thus, the confined states at $\varepsilon_{k}^{\mp}=3JS$ in the LSW Hamiltonian are
linear combinations of type-1 to type-4 building blocks, whereas those at $\varepsilon_{k}=0$
in the tight-binding Hamiltonian are linear combinations of type-1 to type-6 building blocks.
Examples of the coverings of the Penrose lattice by these building blocks are shown in
Figure~\ref{F:covering}.

\section{Boundary-Condition Effects}
   The fractions of type-$n$ building blocks in the thermodynamic limit,
$f_{n}\equiv\lim_{L\to\infty}N_{n}/L$, can be determined exactly using either
the inflation-deflation rule \cite{K6924,A1621} or
the perpendicular-space accounting method \cite{M064213}.
These two independent methods conclude the same results.
The values of $f_{n}$ are summarized in Table~\ref{T:fraction}.
The total fraction of confined states is given by
$\sum_{n=1}^{4}f_{n}=280-173\tau\simeq 8.012\times 10^{-2}$ for the LSW confined states
at $\varepsilon_{k}^{\mp}=3JS$, and by
$\sum_{n=1}^{6}f_{n}=81-50\tau\simeq 9.830\times 10^{-2}$ for the tight-binding confined
states at $\varepsilon_{k}=0$,
where $\tau\equiv\frac{1+\sqrt{5}}{2}$ is the golden number.

\begin{table}
\caption{%
         The fraction of each building block.
         }
\centering
\begin{ruledtabular}
\begin{tabular}{ll}
Building blocks & Their fraction
\\
\hline
Type-1 & $f_{1}=\cfrac{1}{\tau^{8}(1+\tau^{2})} \simeq 5.883\times 10^{-3}$
\vspace{1mm}
\\
Type-2 & $f_{2}=\cfrac{1}{\tau^{6}(1+\tau^{2})} \simeq 1.540\times 10^{-2}$
\vspace{1mm}
\\
Type-3 & $f_{3}=\cfrac{2}{\tau^{8}} \simeq 4.257 \times 10^{-2}$
\vspace{1mm}
\\
Type-4 & $f_{4}=\cfrac{2}{\tau^{10}} \simeq 1.626 \times 10^{-2}$
\vspace{1mm}
\\
Type-5 & $f_{5}=\cfrac{1}{\tau^{9}} \simeq 1.316\times 10^{-2}$
\vspace{1mm}
\\
Type-6 & $f_{6}=\cfrac{1}{\tau^{11}} \simeq 5.025\times 10^{-3}$
\vspace{1mm}
\\
\hline
Total (LSW) & $\sum\limits_{n=1}^{4}f_{n}=280-173\tau\simeq 8.012\times 10^{-2}$
\vspace{1mm}
\\
\hline
Total (tight-binding) & $\sum\limits_{n=1}^{6}f_{n}=81-50\tau\simeq 9.830\times 10^{-2}$
\end{tabular}
\end{ruledtabular}
\label{T:fraction}
\end{table}

\begin{figure}
\centering
\includegraphics[width=\linewidth]{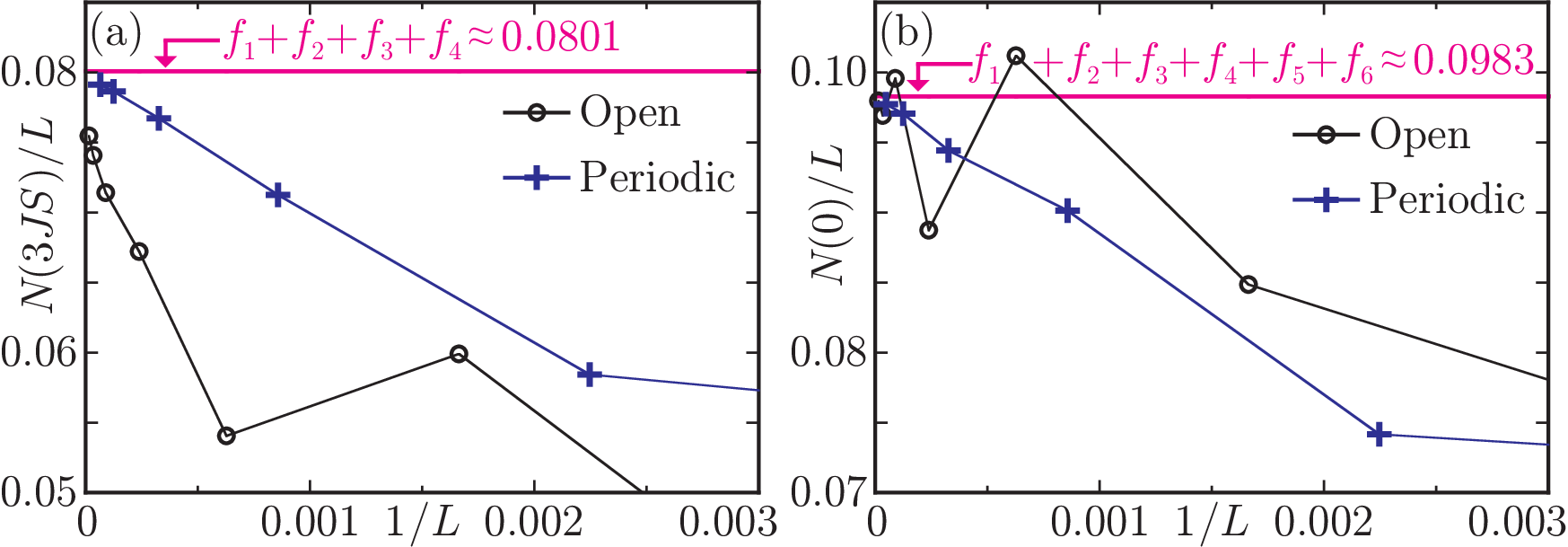}
\caption{%
         Finite-size scaling of the fraction of confined states at
         $\varepsilon_{k}^{\mp}=3JS$ for the LSW Hamiltonian (a) and at
         $\varepsilon_{k}=0$ for the tight-binding Hamiltonian (b),
         plotted as a function of $1/L$.
         Black circles represent results for open-boundary clusters,
         while blue crosses correspond to boundary-free periodic approximants.
         Magenta lines denote the analytically estimated fractions
         in the thermodynamic limit: 
         $\sum_{n=1}^{4}f_{n}=280-173\tau$ for the LSW Hamiltonian
         and $\sum_{n=1}^{6}f_{n}=81-50\tau$ for the tight-binding Hamiltonian.
         }
\label{F:fractionFS}
\end{figure}

   Let $N(\hbar\omega)$ denote the number of states with energy $\hbar\omega$.
Figure~\ref{F:fractionFS} shows the finite-size scaling of the fractions $N(3JS)/L$ and
$N(0)/L$ obtained numerically, corresponding to the $\varepsilon_{k}^{\mp}=3JS$ eigenstates
of the LSW Hamiltonian and  $\varepsilon_{k}=0$ eigenstates of the tight-binding Hamiltonian,
respectively.
For finite $L$, the fraction of confined states deviates from its thermodynamic-limit value.
Since the wavefunctions of the confined states are strictly localized within the finite-extent
building blocks shown in Figure~\ref{F:buildingblocks}, the Penrose lattice must be
sufficiently large to accommodate a proper tiling by these building blocks.
In the thermodynamic limit, the fraction of LSW $\varepsilon_{k}^{\mp}=3JS$ confined
states $N(3JS)/L$ converges to the estimated value  $280-173\tau$ in  boundary-free
periodic approximants, but remains approximately $6\%$ lower even at $L=65751$ in an
open-boundary cluster, as is shown in Figure~\ref{F:fractionFS}a.
In contrast, as shown in Figure~\ref{F:fractionFS}b, the fraction of tight-binding
$\varepsilon_{k}=0$ confined states $N(0)/L$ converges to the exact value  $81-50\tau$
regardless of the boundary conditions.

\begin{figure}
\centering
\includegraphics[width=0.67\linewidth]{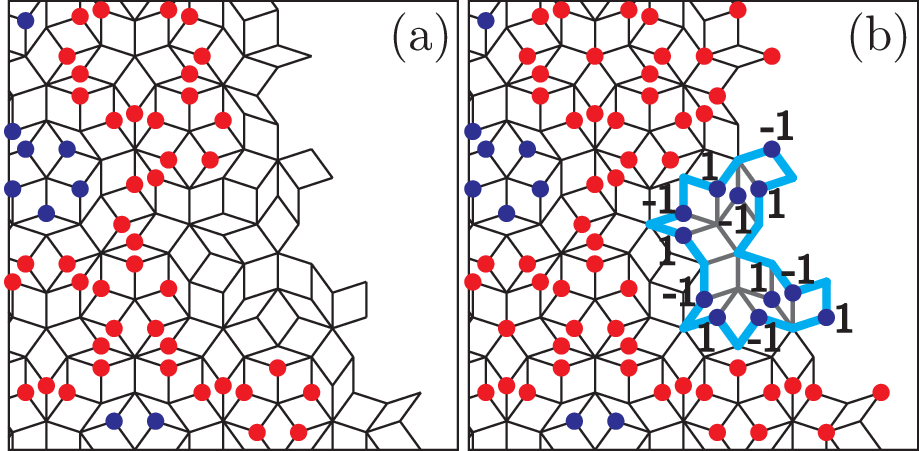}
\caption{%
         Effects of system boundaries on the confined states.
         Partial views of the confined states at $\varepsilon_{k}^{\mp}=3JS$ for
         the LSW Hamiltonian (a) and at $\varepsilon_{k}=0$ for the tight-binding
         Hamiltonian (b) on the Penrose lattice with $L=1591$ sites.
         Each vertex is marked in the same manner as in Figure~\ref{F:confinedreal}.
         The region outlined in light blue in (b) does not correspond to any of the type-1
         to type-6 building blocks, yet satisfies constraint~\eqref{E:confinedconstraintT}.
         The number adjacent to each dot indicates the relative amplitude of
         the unnormalized wavefunction.
         }
\label{F:boundary}
\end{figure}

   Figure~\ref{F:boundary} illustrates the effects of open boundaries.
For the LSW confined states, blank domains without local solutions emerge near the boundaries,
as shown in Figure~\ref{F:boundary}a.
This arises because the coordination number $z_{l}$ at boundary sites is reduced compared
to the bulk, making it virtually impossible to simultaneously satisfy both the constraint
that the wavefunction amplitude is nonzero only on tricoordinated sites and
either Equation~\eqref{E:E3JSconst1} or \eqref{E:E3JSconst2}.
As a result, ribbon-like blank domains, approximately one building block wide, form along
the boundary and lead to an underestimation of $N(3JS)/L$ relative to the exact value.
The fraction of such blank domains is estimated to be $O(1/\sqrt{L})$ in an $L$-site system,
which is consistent in order of magnitude with the observed discrepancy in $N(3JS)/L$.
In contrast, in boundary-free periodic approximants, the coordination numbers remain unaffected,
thus preventing the formation of such blank domains.
Consequently, the fraction $N(3JS)/L$ converges to the exact value $280-173\tau$,
demonstrating that only type-1 to type-4 building blocks constitute the LSW confined states.
In the case of the tight-binding Hamiltonian, i.e., the matrix $\mathbf{T}$,
some of the $\varepsilon_{k}=0$ eigenstates include building blocks other than types 1 to 6
as is shown in Figure~\ref{F:boundary}b.
As a result, the entire Penrose lattice can be covered, excluding the one-dimensional
forbidden ladders, and $N(0)/L$ converges to the exact value $81-50\tau$ under both boundary
conditions.

   Boundary-induced reduction of $N(3JS)$ may likewise occur in other quasiperiodic systems with
confined states constructed from particular local building blocks, each consisting of
same-coordinated sites.
We evaluate the fraction of the confined states for several open clusters of various sizes,
which effectively correspond to different boundary terminations.
The calculated fraction converges to the analytically derived value for the infinite-size lattice.
This suggests that changing the boundary termination does not qualitatively alter the formation
of finite-width blank domains near the boundaries.
In addition, matching-rule-violating defect sites (see Appendix~\ref{S:periodicapprox}
and Figure~\ref{F:periodicapprox}) emerge in periodic approximants of the Penrose lattice
and locally obstruct the placement of the building blocks.
As a result, point-like blank domains are formed around such defect sites, whereas boundary
effects produce one-dimensional blank domains.
The affected fractions therefore scale as $O(1/L)$ and $O(1/\sqrt{L})$, respectively.
Owing to the local building-block nature of the LSW confined states on the Penrose lattice,
the effects of boundaries, defects, and local disorder on the bulk confined-state fraction
are expected to vanish in the thermodynamic limit.

\section{Superconfined Antiferromagnons}
\begin{figure}
\centering
\includegraphics[width=\linewidth]{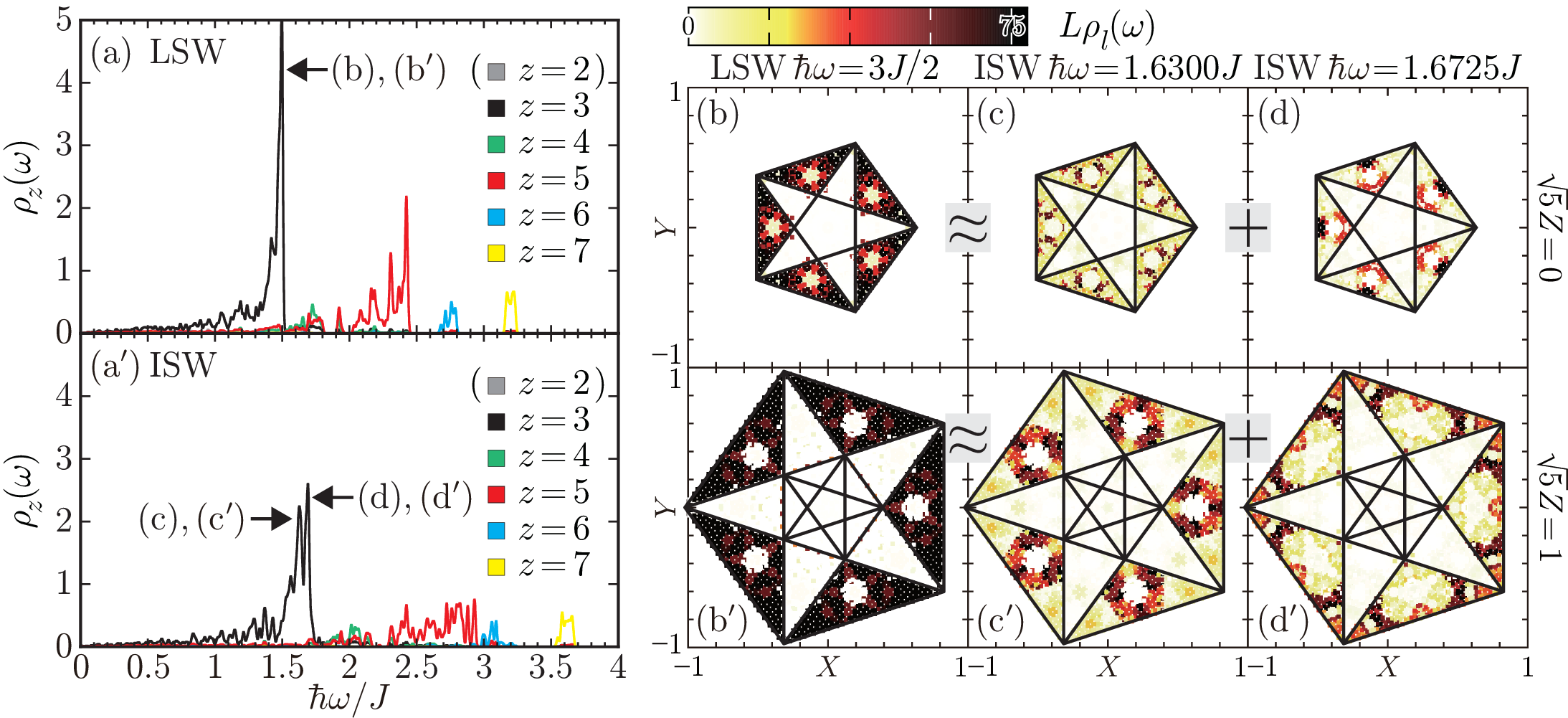}
\caption{%
         (Left panels)
         The coordination-number-resolved density of states $\rho_{z}(\omega)$ for the LSW (a)
         and ISW (a${}^{\prime}$) excitation spectra of the $S=\frac{1}{2}$ Heisenberg model
         on the Penrose lattice of $L=11006$.
         Arrows in (a) and (a${}^{\prime}$) indicate the energies at which
         the perpendicular-space analyses of the site-resolved density of states
         are conducted.
         (Right panels)
         Contour plots in the perpendicular space of the site-resolved density of states
         $\rho_{l}(\omega)$ multiplied by the number of sites $L$ for the LSW spectrum
         at $\hbar\omega=3J/2$ [(b), (b${}^{\prime}$)], and for the ISW spectrum
         at $\hbar\omega=1.6300J$ [(c), (c${}^{\prime}$)] and $\hbar\omega=1.6725J$ 
         [(d), (d${}^{\prime}$)].
         While results for $\sqrt{5}Z=0$ and $\sqrt{5}Z=1$ are presented here, 
         the corresponding results for $\sqrt{5}Z=3$ and $\sqrt{5}Z=2$ are equivalent,
         respectively.
         }
\label{F:OS0correction}
\end{figure}

   We incorporate the $O(S^{0})$ quantum corrections (see Appendix~\ref{S:ISW}) into the
LSW Hamiltonian via the Wick decomposition \cite{N034714,Y094412}.
Figure~\ref{F:OS0correction} compares the site-resolved density of states for the LSW and
up-to-$O(S^{0})$ interacting spin-wave (ISW) excitation spectra at $S=\frac{1}{2}$,
together with their mappings in the perpendicular space.
In the LSW spectrum, a prominent peak at $\hbar\omega=3J/2$ corresponds to
massively degenerate confined states (Figure~\ref{F:OS0correction}a).
The $O(S^{0})$ magnon-magnon interactions introduce environment-dependent corrections
into the ISW Hamiltonian \eqref{E:ISWHammatrix}.
In the absence of translational invariance, interaction-induced on-site potentials vary even
among sites with the same coordination number \cite{Y702}. 
As a result, the $O(S^{0})$ quantum corrections not only shift the excitation spectrum
upward but also lift the degeneracy at $\hbar\omega=3J/2$.
The corresponding peaks remain prominent in the ISW spectrum, but split into two peaks
around $1.6300J$ and $1.6725J$ (Figure~\ref{F:OS0correction}a${}^{\prime}$).

   The one-to-one correspondence between positions in the perpendicular space and local 
vertices in the physical space (Figure~\ref{F:perpspace}) confirms that the LSW spectral weight
at $\hbar\omega=3J/2$ (Figures~\ref{F:OS0correction}b and \ref{F:OS0correction}b${}^{\prime}$)
is  localized within the $z=3$ domains.
In these domains, certain regions exhibit negligible spectral intensity, consistent with
the perpendicular-space analysis of forbidden regions \cite{M064213} associated with the
type-1 to type-4 building blocks.
The perpendicular-space mappings of the ISW spectrum reveal that the spectral weight remains
concentrated in the $z=3$ domains for both the lower branch at $\hbar\omega=1.6300J$
(Figures~\ref{F:OS0correction}c and \ref{F:OS0correction}c${}^{\prime}$) and the upper
branch at $\hbar\omega=1.6725J$
(Figures~\ref{F:OS0correction}d and \ref{F:OS0correction}d${}^{\prime}$).
Compared to the LSW case, however, the intensity exhibits a more fragmented distribution,
with each branch occupying distinct subdomains.
The subdivision of the perpendicular-space domains reflects a finer classification of local
environments in the physical space, arising from the inclusion of more distant surrounding
geometry \cite{Y702,G224201}.
The ISW spectral weight distributions of the two branches are complementary within the $z=3$
domains.
This indicates that the $O(S^{0})$ quantum fluctuations distinguish these subtle geometric
variations and resolve subsets of tricoordinated sites into distinct energy branches.
The magnons forming these two branches remain well confined to tricoordinated
sites and are thus ``superconfined'' to selected subsets of these sites.

\section{Summary and Discussion}
   In the vertex model on the Penrose lattice, itinerant electrons exhibit
macroscopically degenerate confined states at $\varepsilon_{k}=0$.
They consist of type-1 to type-6 building blocks.
In contrast, in the antiferromagnetic Heisenberg model on the Penrose lattice,
localized spins exhibit
macroscopically degenerate confined states at $\varepsilon_{k}^\mp=3JS$ within
the harmonic-oscillator approximation.
They consist only of type-1 to type-4 building blocks.
The spin-wave Hamiltonian in the site representation has coordination-number-dependent
positive on-site potentials, which more stabilize lower-coordinated sites.
Type-1 to type-4 building blocks consist of tricoordinated sites only, while
type-5 and type-6 building blocks consist of pentacoordinated as well as tricoordinated
sites.
Confined spin waves consist purely of tricoordinated sites.

\begin{figure}
\centering
\includegraphics[width=\linewidth]{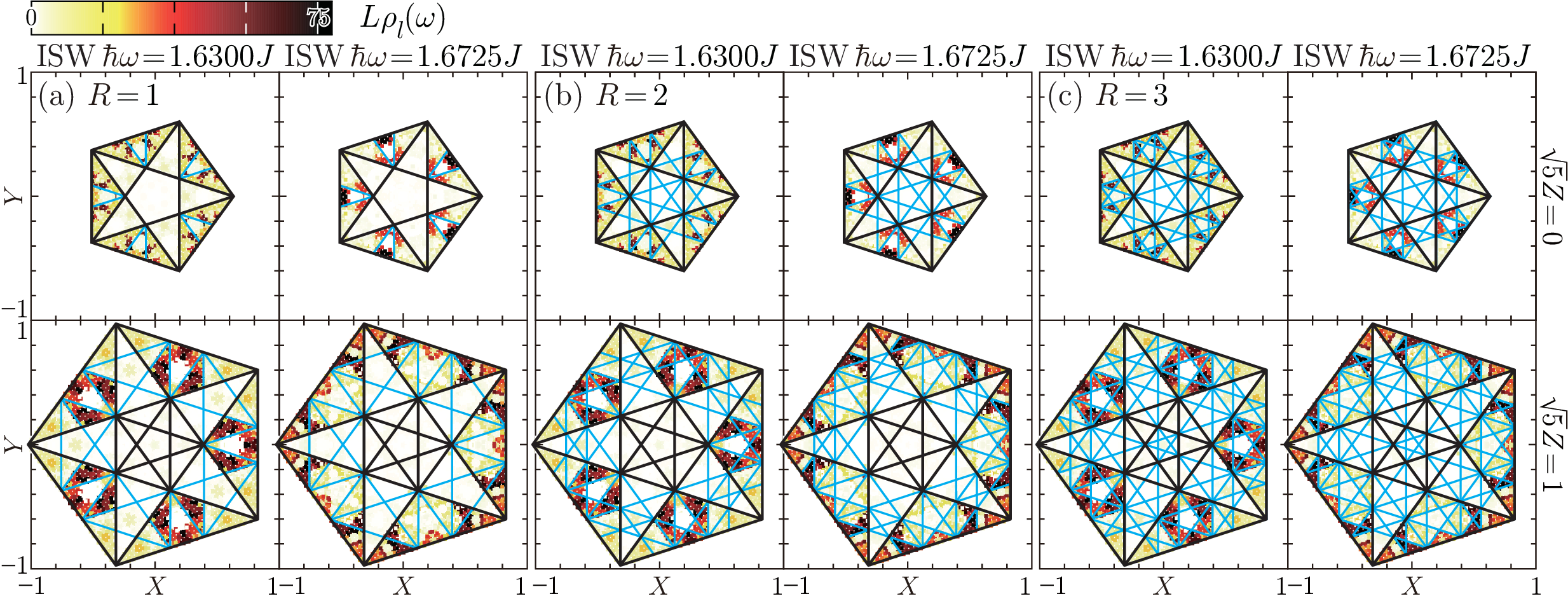}
\caption{%
         Contour plots of the site-resolved density of states $L\rho_{l}(\omega)$ in the
         perpendicular space for the ISW spectrum at $\hbar\omega=1.6300J$ and $1.6725J$,
         overlaid with lines indicating the subdivision of the perpendicular-space domains.
         The domains are subdivided according to their surrounding
         environments in the physical space, defined by all vertices within a distance $R$
         from each vertex.
         Initially corresponding to the 8 types of vertices of the Penrose lattice (black lines),
         these domains are further subdivided into 15, 27, and 40 types for $R=1$ (a),
         $R=2$ (b), and $R=3$ (c), respectively (cyan lines).
         }
\label{F:perpsubclassRdep}
\end{figure}

   What will happen to confined antiferromagnons when they are brought into interaction?
The $O(S^{0})$ quartic interactions lift the macroscopic degeneracy at
$\varepsilon_{k}^{\mp}=3JS$ to yield two branches, as is shown in Figure~\ref{F:OS0correction}.
Their spectral weights are distributed complementarily in the perpendicular space.
In order to formulate these distributions, we define a distance $R$ between two vertices
such that the shortest route from one to the other consists of $R$ bonds.
With increasing $R$, vertices are more and more classified into subgroups and subdomains
in the physical and perpendicular spaces, respectively.
The eight fundamental domains are further subdivided into 15, 27, and 40 types for $R=1$,
$2$, and $3$, respectively.
Overlaying these subdivisions onto the perpendicular-space distributions for each branch shows
that the spectral-weight patterns are well described at $R=3$ (Figure~\ref{F:perpsubclassRdep}).
This can be understood as follows.
As an example, consider an arbitrary bond $\langle i,j \rangle$.
In the LSW Hamiltonian, the matrix elements consist of on-site potentials determined by
the coordination number of each site and uniform nearest-neighbor bonds.
Thus, the LSW description is strictly local to each bond.
In the ISW Hamiltonian, the $O(S^{0})$ corrections to the quadratic terms contain site
expectation values $\langle a_{i}^{\dagger}a_{i} \rangle$ and
$\langle b_{j}^{\dagger}b_{j} \rangle$, and bond expectation values $\langle a_{i}b_{j} \rangle$
and $\langle a_{i}^{\dagger}b_{j}^{\dagger} \rangle$, as is shown in
Equation~\eqref{E:H0matrixelement} (see Appendix~\ref{S:ISW}).
These quantities depend on their local environments and therefore vary from bond to bond.
They are self-consistently determined via matrix elements depending on the surrounding
environments of each bond.
Thus, the ISW description includes contributions beyond the bond in issue.
Specifically, for the sites $\bm{r}_{i}$ and $\bm{r}_{j}(=\bm{r}_{i}+\bm{\delta})$,
the neighboring sites $\bm{r}_{j'}=\bm{r}_{i}+\bm{\delta}_{i}$ ($\neq \bm{r}_{j}$) and
$\bm{r}_{i'}=\bm{r}_{j}+\bm{\delta}_{j}$ ($\neq \bm{r}_{i}$) also explicitly enter the
ISW Hamiltonian.
These sites are related by $\bm{r}_{i'}=\bm{r}_{j'}-\bm{\delta}_i+\bm{\delta}+\bm{\delta}_j$,
implying that contributions up to a distance $R=3$ are relevant at the $O(S^{0})$ level.
This estimate is consistent with the spectral-weight distributions of the superconfined
antiferromagnons.

   What will happen to superconfined antiferromagnons when they are brought into
further interactions higher than $O(S^0)$?
The $O(S^{-1})$ sextic interactions and still higher-order terms introduce additional
corrections to the eigenvalues.
Since the nearest-neighbor exchange interaction conserves magnetization, the expectation values
in the quadratic terms obtained via Wick decomposition remain limited to the same types as
in the $O(S^{0})$ case: site expectation values $\langle a_{i}^{\dagger}a_{i} \rangle$ and
$\langle b_{j}^{\dagger}b_{j} \rangle$, and bond expectation values $\langle a_{i}b_{j} \rangle$
and $\langle a_{i}^{\dagger}b_{j}^{\dagger} \rangle$.
These expectation values are defined only on sites and nearest-neighbor bonds.
The spatial range of the corrections remains unchanged from the $O(S^{0})$ case.
Thus, such higher-order corrections are unlikely to incorporate more distant environments
to induce further branching.
\textit{Superconfinement} is not fractionalization up to infinity but dividing
tricoordinated sites into essentially two groups.

%experiment
Resonant inelastic x-ray scattering (RIXS) \cite{H167404,M021041,BL012053}
may be a functional light probe to capture confined antiferromagnons of our interest.
While spin-orbit-coupling-assisted magnetic Raman scattering \cite{F514,Ixxxx} serves to
reveal single magnon eigenlevels, the Raman operator of this type reads a total uniform
magnetization in the polarization-dependent particular direction and therefore acts only
on rotation-invariant and mirror-symmetric magnons.
It is thus useless to detect confined antiferromagnons consisting of mirror-antisymmetric
building blocks.
The single-magnon Raman scattering is further disappointing because it cannot have any access
to the $\mathrm{SU}(2)$ symmetric Heisenberg Hamiltonian commutable with the total uniform
magnetization in every direction.
Inelastic neutron scattering \cite{Y702,W104427} is indeed available in this context,
but it requires relatively large bulk samples.
RIXS works even with small samples and thin films.
We can tune the momentum, polarization, and resonance energy of incoming and outgoing x-rays.
Interestingly enough, photons can distinguish between sublattices.
Circularly polarized light interacts with either a magnetization-enhancing or
magnetization-reducing magnon.
Each confined state resides in either A or B sublattice, because its building blocks belong to
either A or B sublattice.
There is a possibility of selectively accessing confined antiferromagnons
according to their constituent building blocks.
RIXS spectra possibly contain a nearly flat band at $\varepsilon_{k}^{\mp}\simeq 3JS$.

%%%%%%%%%%%%%%%%%%%%%%%%%%%%%%%%%%%%%%%%%%
\vspace{6pt}

%%%%%%%%%%%%%%%%%%%%%%%%%%%%%%%%%%%%%%%%%%
%\authorcontributions{
%Conceptualization, methodology, validation, formal analysis, investigation, T.I. and S.Y.; resources, S.Y.; writing manuscript, T.I. and S.Y.; supervision, project administration, funding acquisition, S.Y. All authors have read and agreed to the published version of the manuscript.}

%\funding{This work is supported by JSPS KAKENHI Grant Number 22K03502.}
%\funding{This work is supported by ... ...}

%\institutionalreview{Not applicable}

%\dataavailability{Any data that support the findings of this study are included within the article.}

%\acknowledgments{The authors are grateful to A. Koga for his useful comments on our argument.}

%\conflictsofinterest{The authors declare no conflicts of interest.} 
%%%%%%%%%%%%%%%%%%%%%%%%%%%%%%%%%%%%%%%%%%

\appendix
%%\section[\appendixname~\thesection]{Penrose Lattice}
\section{Penrose Lattice}
\label{S:PenroseLattice}
%%\subsection[\appendixname~\thesubsection]{Geometric Properties}
\subsection{Geometric Properties}
\label{S:PenroseGeometry}
\begin{figure}[b] %[htb]%[H]
\centering
\includegraphics[width=\linewidth]{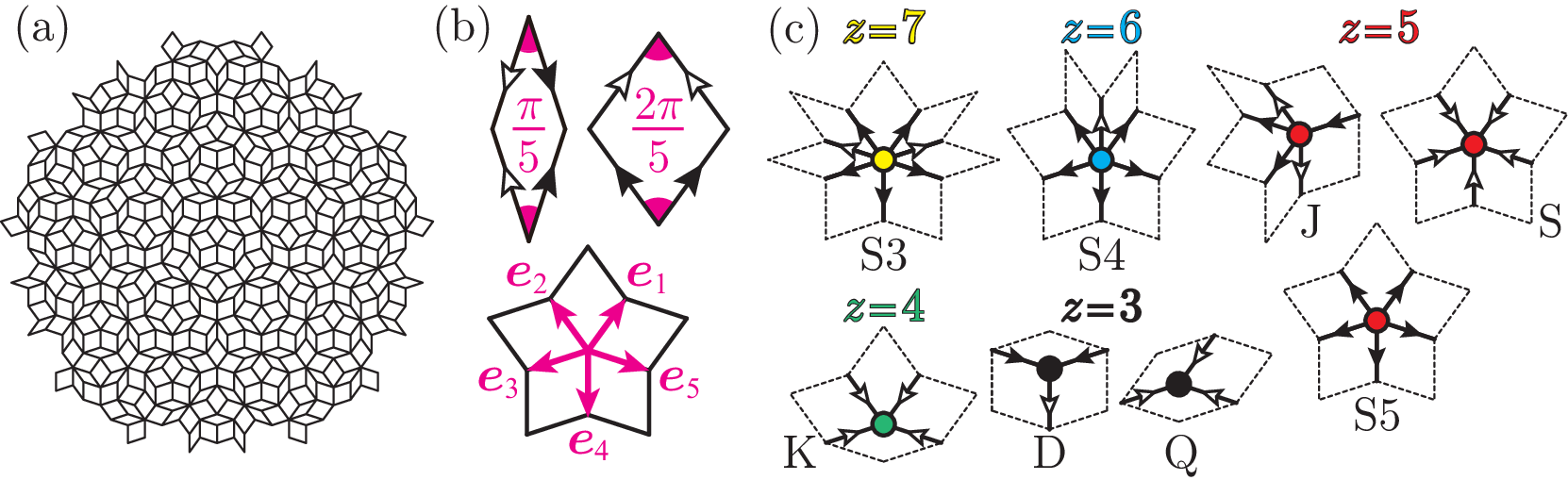}
\caption{%
         (a) A two-dimensional Penrose lattice of size $L=526$ with fivefold rotational symmetry.
         (b) The Penrose lattice consists of two rhombuses with acute angles $\frac{\pi}{5}$
             and $\frac{2\pi}{5}$, respectively.
             Each edge is marked with an arrow to define the matching rules.
             The canonical basis vectors of a five-dimensional hypercubic lattice are
             projected onto five vectors, denoted by
             $\bm{e}_{1}$, $\bm{e}_{2}$, $\bm{e}_{3}$, $\bm{e}_{4}$, and $\bm{e}_{5}$.
             Any four of these vectors can be chosen as the primitive translation vectors
             for the two-dimensional Penrose lattice.
             Note that $\sum_{n=1}^{5}\bm{e}_{n}=\bm{0}$.
         (c) Eight types of vertices in the Penrose lattice with 
             coordination numbers $z$ ranging from $3$ to $7$ \cite{dB39}.
         }
\label{F:PenroseLattice}
\end{figure}

   A two-dimensional Penrose lattice (Figure~\ref{F:PenroseLattice}a) can be generated from
two types of rhombuses with acute angles of $\frac{\pi}{5}$ and $\frac{2\pi}{5}$
(Figure~\ref{F:PenroseLattice}b) by matching edges with identical arrow markings and
iteratively applying the inflation-deflation operation \cite{S3904,K6924}.
These operations yield self-similarity with a magnification ratio equal to the golden number
$\tau=\frac{1+\sqrt{5}}{2}$ \cite{G110}.
Although the Penrose lattice has a crystallographically forbidden fivefold rotational symmetry,
its diffraction pattern consists of sharp $\delta$-function peaks \cite{M609}
and exhibits long-range positional order.
While the physical space dimension $d$ of the Penrose lattice is $2$,
its indexing dimension (or simply rank) $D$ is $4$.
Every vertex of the Penrose lattice is represented by an integer linear combination of
four independent primitive lattice vectors (Figure~\ref{F:PenroseLattice}b).
The Penrose lattice is bipartite, consisting of two sublattices, which we refer to as A and B.
In general, a $d$-dimensional quasiperiodic lattice can be obtained from
a higher-than-$d$-dimensional periodic lattice via an algebraic approach \cite{dB39,dB53}.
Both the Penrose lattice and its periodic approximants can be obtained as the projection of
a five-dimensional hypercubic lattice onto a two-dimensional physical space
(see Appendix~\ref{S:periodicapprox}).
The infinite Penrose lattice consists of eight types of local vertices whose coordination
numbers $z$ range from $3$ to $7$, as is shown in Figure~\ref{F:PenroseLattice}c \cite{dB39}.

\begin{figure} %[H]
\centering
\includegraphics[width=\linewidth]{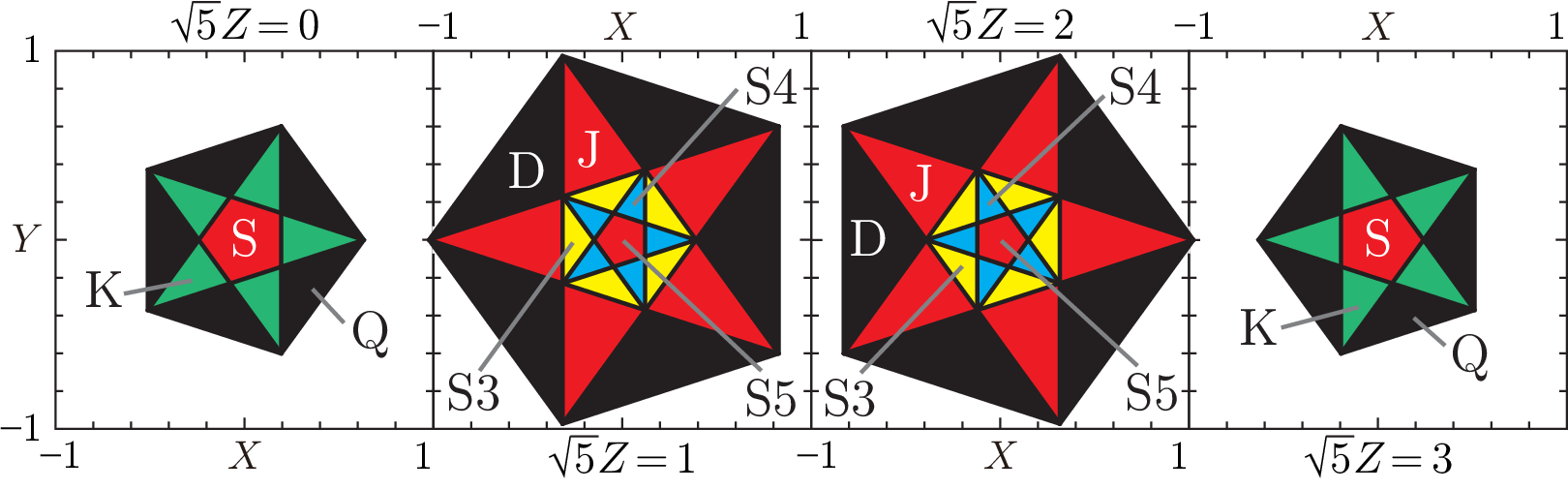}
\caption{%
         The perpendicular space of the Penrose lattice consists of a three-dimensional
         stack of four pentagons lying at $\sqrt{5}Z=0, 1, 2, 3$.
         Each pentagon is divided into several color-coded domains, each of which contains
         a single species of vertices and is labeled with one of the eight types, D to S3,
         as is defined in Figure~\ref{F:PenroseLattice}c.
         }
\label{F:perpspace}
\end{figure}

   The orthogonal complement of the two-dimensional Penrose lattice is a three-dimensional
stack of four pentagons (Figure~\ref{F:perpspace}).
The layers at $\sqrt{5}Z=0, 2$ correspond to the A sublattice, while those at
$\sqrt{5}Z=1,3$ correspond to the B sublattice [cf. Equation~\eqref{E:projectionperp}].
Each of the eight vertex types maps onto a distinct compact region within the
pentagons in the perpendicular space.

%\subsection[\appendixname~\thesubsection]{Projection Method and Periodic Approximants of the Penrose Lattice}
\subsection{Projection Method and Periodic Approximants of the Penrose Lattice}
\label{S:periodicapprox}
   The Penrose lattice is constructed by projecting a five-dimensional hypercubic lattice
onto a two-dimensional physical space \cite{S104427,D2688}.
The projection matrix for the two-dimensional physical-space coordinates $(x,y)$ is
\begin{align}
   &
   \left[
   \begin{array}{c}
     x
    \\
     y
   \end{array}
   \right]
  =
   \sqrt{\frac{2}{5}}
   \left[
   \begin{array}{ccccc}
     1 & \cos\phi_{1} & \cos\phi_{2} & \cos\phi_{3} & \cos\phi_{4}
    \\
     0 & \sin\phi_{1} & \sin\phi_{2} & \sin\phi_{3} & \sin\phi_{4}
   \end{array}
   \right]
   \left[
   \begin{array}{c}
     m_{1}
    \\
     m_{2}
    \\
     m_{3}
    \\
     m_{4}
    \\
     m_{5}
   \end{array}
   \right];
  \ \ 
   \phi_{n}
   \equiv
   \frac{2\pi}{5}n,
  \ 
   m_{n}\in\mathbb{Z}.
\label{E:projectionphysical}
\end{align}
The projection matrix for the three-dimensional perpendicular-space coordinates $(X,Y,Z)$ is
\begin{align}
   \left[
   \begin{array}{c}
     X
    \\
     Y
    \\
     Z
   \end{array}
   \right]
  =
   \sqrt{\frac{2}{5}}
   \left[
   \begin{array}{ccccc}
     1 & c_{2} & c_{4} & c_{6} & c_{8}
    \\
     0 & s_{2} & s_{4} & s_{6} & s_{8}
    \\
     \frac{1}{\sqrt{2}} & \frac{1}{\sqrt{2}} & \frac{1}{\sqrt{2}} & \frac{1}{\sqrt{2}} & \frac{1}{\sqrt{2}}
   \end{array}
   \right]
   \left[
   \begin{array}{c}
     m_{1}
    \\
     m_{2}
    \\
     m_{3}
    \\
     m_{4}
    \\
     m_{5}
   \end{array}
   \right],
\label{E:projectionperp}
\end{align}
where $c_{n}=\cos\phi_{n}$ and $s_{n}=\sin\phi_{n}$ for the quasiperiodic Penrose lattice.
Only the five-dimensional integer vectors ${}^{t}[m_{1}, m_{2}, m_{3}, m_{4}, m_{5}]$ whose
projections onto the three-dimensional perpendicular space fall within a certain region,
called the \textit{selection window}, are retained as lattice points in the two-dimensional
physical space.
The selection window of the Penrose lattice is a rhombic icosahedron, which corresponds to
the projection of a five-dimensional unit hypercube.
Its cross-section, containing the selected lattice points, consists of a stack of four
regular pentagons at positions  where $\sqrt{5}Z\in\mathbb{Z}$,
as is shown in Figure~\ref{F:perpspace}.

\begin{figure}
\centering
\includegraphics[width=0.5\linewidth]{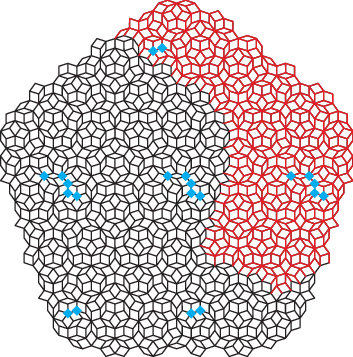}
\caption{%
         Physical-space projection for $-7\leq m_{n} \leq 7$ using the periodic approximant
         projection matrix with Fibonacci index $i=5$ (black),
         along with corresponding periodic unit cell consisting of $L=445$ sites (red).
         Light blue diamonds indicate vertices that violate the local environments
         of the Penrose lattice depicted in Figure~\ref{F:PenroseLattice}c.
         }
\label{F:periodicapprox}
\end{figure}

   In order to construct periodic approximants, $c_{n}$ and $s_{n}$ in
Equation~\eqref{E:projectionperp} are replaced by the following rational approximations:
\begin{align}
   &
   c_{2}=-\frac{\tau_{1}}{2},\ 
   c_{4}=\frac{\tau_{1}-1}{2},\ 
   c_{6}=\frac{\tau_{1}-1}{2},\ 
   c_{8}=-\frac{\tau_{1}}{2},
   \allowdisplaybreaks
   \nonumber \\
   &
   s_{2}=\tau_{2}-1,\ 
   s_{4}=\tau_{3}-\tau_{2}-1,\ 
   s_{6}=1,\ 
   s_{8}=1-\tau_{2},
\label{E:rationalsincos}
\end{align}
where
\begin{align}
   &
   \tau_{1}=\frac{F_{i+2}+F_{i}}{F_{i+1}+F_{i-1}},\ 
   \tau_{2}=1+\frac{F_{i+1}(F_{i}+F_{i-2})}{F_{i-1}(F_{i+3}+F_{i+1})},\ 
   \tau_{3}=\frac{F_{i+4}+F_{i+2}}{F_{i+3}+F_{i+1}}
\label{E:Fibonacciapprox}
\end{align}
and $F_{i}$ denotes the Fibonacci numbers defined by the recurrence relation
$F_{0}=0,\ F_{1}=1,\ F_{i}=F_{i-1}+F_{i-2}\ (i\geq 2)$ \cite{Z3377}.
This procedure yields a periodic unit cell in the two-dimensional physical space that
reproduces the local environments of the Penrose lattice at the majority of vertices,
as is illustrated in Figure~\ref{F:periodicapprox}.
Vertices sharing identical coordinates in the three-dimensional perpendicular space are
connected by translation operations in the two-dimensional physical space.

%\section[\appendixname~\thesection]{Confined States in the Tight-Binding Model on the Penrose Lattice}
\section{Confined States in the Tight-Binding Model on the Penrose Lattice}
\label{S:TBmodel}
%\subsection[\appendixname~\thesubsection]{Tight-Binding Hamiltonian}
\subsection{Tight-Binding Hamiltonian}
\label{S:TBHamiltonian}
   We define the tight-binding Hamiltonian on the two-dimensional Penrose lattice
with $L=L_{\mathrm{A}}+L_{\mathrm{B}}$ sites
\begin{align}
   &
%   \mathscr{H}
   \mathcal{H}_{t}
  =-t\sum_{i\in\mathrm{A}}\sum_{j\in\mathrm{B}}l_{i,j}
   \left(c_{i}^{\dagger}c_{j}+c_{j}^{\dagger}c_{i}\right)
  \equiv
   -t\bm{c}^{\dagger}\mathbf{T}\bm{c}
  \equiv
   -t
   \left[\bm{c}_{\mathrm{A}}^{\dagger}, \bm{c}_{\mathrm{B}}^{\dagger}\right]
   \left[
   \begin{array}{c|c}
     \mathbf{O}_{\mathrm{AA}} & \mathbf{C}_{\mathrm{AB}}
    \\ \hline
     \mathbf{C}_{\mathrm{AB}}^{\dagger} & \mathbf{O}_{\mathrm{BB}}
   \end{array}
   \right]
   \left[
   \begin{array}{c}
     \bm{c}_{\mathrm{A}}
    \\
     \bm{c}_{\mathrm{B}}
   \end{array}
   \right];
   \allowdisplaybreaks
   \nonumber \\
   &
   \bm{c}_{\mathrm{A}}^{\dagger}
  =\left[c_{1}^{\dagger}, \cdots, c_{L_{\mathrm{A}}}^{\dagger}\right],\ 
   \bm{c}_{\mathrm{B}}^{\dagger}
  =\left[c_{L_{\mathrm{A}}+1}^{\dagger},\cdots,c_{L_{\mathrm{A}}+L_{\mathrm{B}}}^{\dagger}\right],
  \ 
   \left[\mathbf{O}_{\mathrm{AA}}\right]_{i,i'}
  =\left[\mathbf{O}_{\mathrm{BB}}\right]_{j,j'}
  =0,
  \ 
   \left[\mathbf{C}_{\mathrm{AB}}\right]_{i,j}=l_{i,j},
\label{E:TBHam}
\end{align}
where the site indices are labeled as $i\in\mathrm{A}$ and $j\in\mathrm{B}$.
We denote a generic site---meaning one that may belong to either the A or B sublattice---by
$\bm{r}_{l}$, and let $c_{l}^{\dagger}$ denote the creation operator of an electron at site
$\bm{r}_{l}$.
Note that we omit the spin degrees of freedom of the electrons, as they do not affect 
the present discussion.
$\bm{c}^{\dagger}$, $\bm{c}_{\mathrm{A}}^{\dagger}$, and $\bm{c}_{\mathrm{B}}^{\dagger}$
are row vectors of dimension $L$, $L_{\mathrm{A}}$, and $L_{\mathrm{B}}$, respectively.
$\mathbf{O}_{\mathrm{AA}}$ and $\mathbf{O}_{\mathrm{BB}}$
are zero matrices of dimension $L_{\mathrm{A}}\times L_{\mathrm{A}}$ and 
$L_{\mathrm{B}}\times L_{\mathrm{B}}$, respectively,
and $\mathbf{C}_{\mathrm{AB}}$ is the biadjacency  %%connectivity
matrix of dimension $L_{\mathrm{A}}\times L_{\mathrm{B}}$.
The linkage identifier $l_{i,j}$ equals $1$ if the vertices $\bm{r}_{i}$ and $\bm{r}_{j}$
are connected, and $0$ otherwise.
$t$ denotes the transfer integral for electron hopping.
Applying a standard unitary transformation
\begin{align}
   \left[
   \begin{array}{c}
     \bm{c}_{\mathrm{A}}
    \\
     \bm{c}_{\mathrm{B}}
   \end{array}
   \right]
  =
   \mathbf{U}\bm{\alpha};
\ 
   [\mathbf{U}]_{l,k}
  \equiv 
   u_{l,k},
\ 
   \bm{\alpha}^{\dagger}
  \equiv
   \left[\alpha_{1}^{\dagger}, \cdots, \alpha_{L}^{\dagger}\right]
\label{E:unitarytrans}
\end{align}
diagonalizes the Hamiltonian~\eqref{E:TBHam} as
\begin{align}
%   \mathscr{H}
   \mathcal{H}_{t}
  =\sum_{k=1}^{L}\varepsilon_{k}\alpha_{k}^{\dagger}\alpha_{k},
\label{E:TBdiagHam}
\end{align}
where $\alpha_{k}^{\dagger}$ creates a quasiparticle fermion of energy $\varepsilon_{k}$.

\begin{figure}
\centering
\includegraphics[width=0.5\linewidth]{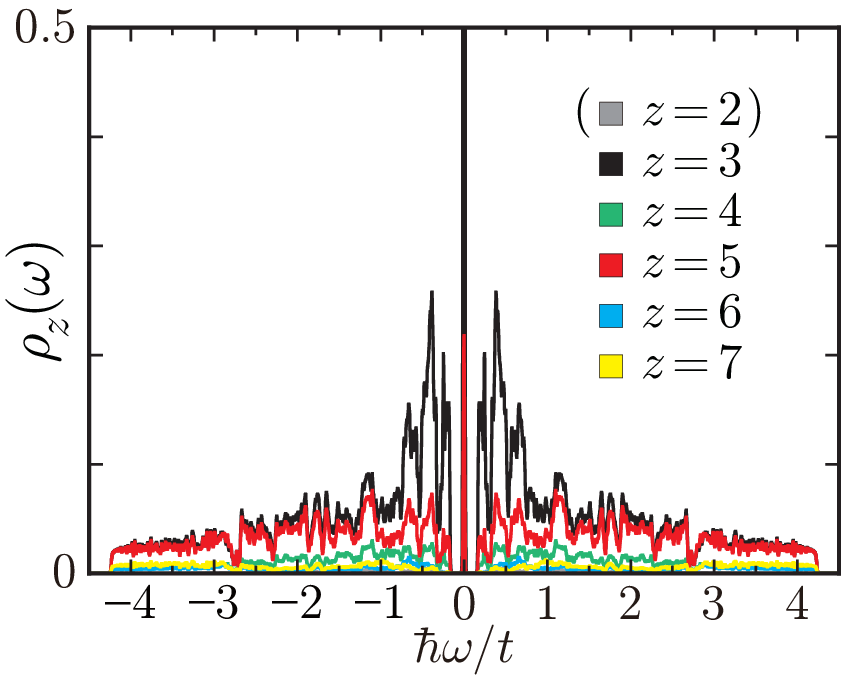}
\caption{%
         The coordination-number-resolved density of states $\rho_{z}(\omega)$ for
         the tight-binding excitation spectrum on the Penrose lattice with $L=11006$,
         where $z$ ranges from $3$ to $7$.
         Note that $z=2$ vertices are artifacts of finite open-boundary clusters.
         }
\label{F:tightbindingLDOS}
\end{figure}

   We define the site-resolved density of states
for the quasiparticle fermion excitation spectrum
\begin{align}
   &
   \rho(\omega)
  =\sum_{l=1}^{L}\rho_{l}(\omega)
  =\sum_{z=3}^{7}\rho_{z}(\omega)
  =\sum_{z=3}^{7}
   \frac{1}{L}\sum_{k=1}^{L}
   \frac{\sum_{l(z_{l}=z)}|u_{l,k}|^{2}}{\sum_{l=1}^{L}|u_{l,k}|^{2}}
   \delta\left(\hbar\omega-\varepsilon_{k}\right),
   \label{E:TBDOS}
\end{align}
where $z_{l}$ is the coordination number of the vertex at $\bm{r}_{l}$.
Figure~\ref{F:tightbindingLDOS} shows the coordination-number-resolved density of states
for the tight-binding excitation spectrum~\eqref{E:TBDOS}.
The density of states is highly singular, exhibiting spikes across all energies.
We find that the $\delta$-function peak at $\hbar\omega=0$ accounts for approximately $10$\%
of the total number of states, which implies the existence of confined states.
This peak is separated from the rest of the spectrum---two symmetric continua---by a gap
of about $0.17t$ \cite{A1621,K214402}.
The $\delta$-function peak is essentially composed of tricoordinated and pentacoordinated
sites, whereas the remaining states have nonzero weights for all coordination numbers.

%\subsection[\appendixname~\thesubsection]{Zero-Energy Confined States}
\subsection{Zero-Energy Confined States}
\label{S:TBconfined}
   We derive the constraints on the zero-energy eigenstates of the connectivity matrix
$\mathbf{T}$.
The eigenvector $\bm{u}_{k}$ corresponding to eigenstate index $k$ is given by
\begin{align}
   &
   \bm{u}_{k}
  =\left[
   \begin{array}{c}
     u_{1,k}
    \\
     \vdots
    \\
     u_{L_{\mathrm{A}},k}
    \\
     u_{L_{\mathrm{A}}+1,k}
    \\
     \vdots
    \\
     u_{L_{\mathrm{A}}+L_{\mathrm{B}},k}
   \end{array}
   \right]
  \equiv
   \left[
   \begin{array}{c}
     \bm{u}_{\mathrm{A};k}
    \\
     \bm{u}_{\mathrm{B};k}
   \end{array}
   \right];
  \ 
   \left[\bm{u}_{\mathrm{A};k}\right]_{i}
  =u_{i,k},
   \ 
   \left[\bm{u}_{\mathrm{B};k}\right]_{j}
  =u_{L_{\mathrm{A}}+j,k}.
\label{E:eigenvectorT}
\end{align}
The eigenvector $\bm{u}_{k}$ satisfies the eigenvalue equation
\begin{align}
   \mathbf{T}\bm{u}_{k}
  =-\frac{\varepsilon_{k}}{t}\bm{u}_{k}.
\label{E:eigenEqT}
\end{align}
Since the connectivity matrix $\mathbf{T}$ is expressed in terms of the biadjacency matrix
$\mathbf{C}_{\mathrm{AB}}$ as shown in Equation~\eqref{E:LSWdiagoffdiag}, the left-hand side of
Equation~\eqref{E:eigenEqT} can be written as
\begin{align}
   &
   \mathbf{T}\bm{u}_{k}
  =\left[
   \begin{array}{c}
     \mathbf{C}_{\mathrm{AB}}\bm{u}_{\mathrm{B};k}
    \\
     \mathbf{C}_{\mathrm{AB}}^{\dagger}\bm{u}_{\mathrm{A};k}
   \end{array}
   \right];
   \ 
   [\mathbf{C}_{\mathrm{AB}}\bm{u}_{\mathrm{B};k}]_{i}
  =\sum_{j\in\mathrm{B}}l_{i,j}u_{L_{\mathrm{A}}+j,k},
  \ 
   [\mathbf{C}_{\mathrm{AB}}^{\dagger}\bm{u}_{\mathrm{A};k}]_{j}
  =\sum_{i\in\mathrm{A}}l_{i,j}u_{i,k}.
\label{E:eigenEqTleft}
\end{align}
Therefore, for $\varepsilon_{k}=0$, the wavefunction amplitudes must satisfy
\begin{align}
   \begin{cases}
     \sum\limits_{j\in\mathrm{B}}l_{i,j}u_{L_{\mathrm{A}}+j,k}=0 &
     (l=i\in\mathrm{A})
    \\
     \sum\limits_{i\in\mathrm{A}}l_{i,j}u_{i,k}=0 &
     (l=j\in\mathrm{B})
   \end{cases}.
\label{E:confinedconstraintT}
\end{align}

   The confined states of the Penrose lattice are divided into discrete domains by the forbidden
ladders (Figure~\ref{F:confinedreal}), which allows each domain to be treated independently.
Within each domain, the confined states have finite amplitudes on only one of the two
sublattices, A or B.
Let us consider a domain in which the amplitudes are finite only on the A-sublattice.
The analogous argument applies to domains with finite amplitudes only on the B-sublattice.
Equation~\eqref{E:confinedconstraintT} is automatically satisfied for all A-sublattice sites
$\bm{r}_{i}$, since $u_{L_{\mathrm{A}}+j,k}=0$ for all adjacent B-sublattice sites $\bm{r}_{j}$.
In contrast, for B-sublattice sites within the same domain, Equation~\eqref{E:confinedconstraintT}
must be satisfied explicitly, since $u_{i,k}$ may be nonzero.
There are exactly six types of local solutions satisfying these constraints in the infinite Penrose lattice, as is shown in Figure~\ref{F:buildingblocks} \cite{K2740,A1621,K214402}.
The $\varepsilon_{k}=0$ eigenstates of the tight-binding Hamiltonian are linear combinations
of type-1 to type-6 building blocks on the Penrose lattice.

%\subsection[\appendixname~\thesubsection]{Coverings of the Penrose Lattice Using the Building Blocks of Confined States}
\subsection{Coverings of the Penrose Lattice Using the Building Blocks of Confined States}
\label{S:covering}
\begin{figure}%[H]
\centering
\includegraphics[width=\linewidth]{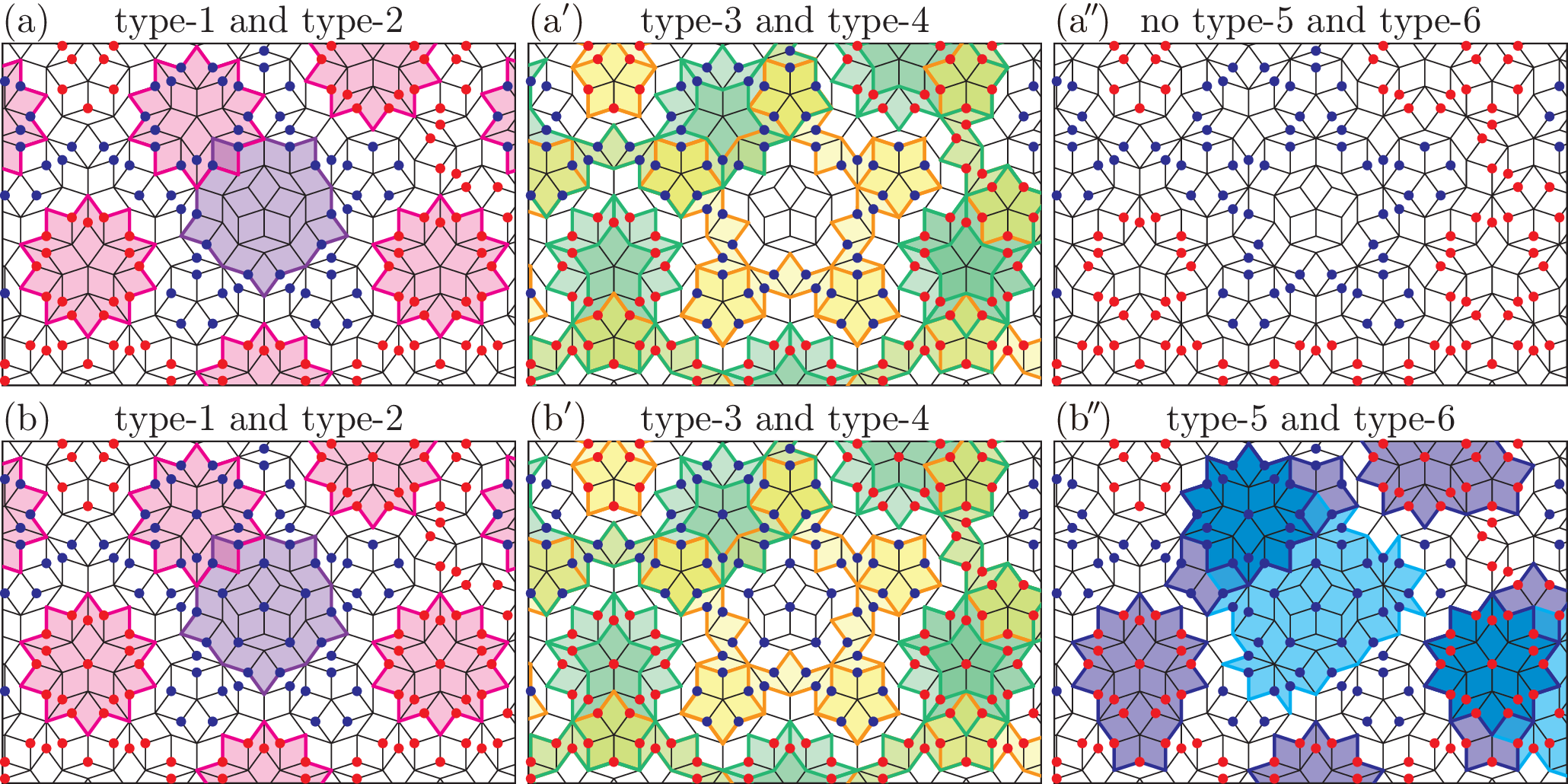}
\caption{%
         Enlarged views of the rectangular regions highlighted in
         Figure~\ref{F:confinedreal}.
         The $\varepsilon_{k}^{\mp}=3JS$ eigenstates of the LSW Hamiltonian
         [(a)-(a${}^{\prime\prime}$)]
         and $\varepsilon_{k}=0$ eigenstates of the tight-binding Hamiltonian
         [(b)-(b${}^{\prime\prime}$)].
         Color-coded regions indicate the building blocks of each confined state
         (magenta: type-1,
          purple: type-2,
          yellow: type-3,
          green: type-4,
          blue: type-5, and
          light blue: type-6).
         The actual confined states at $\varepsilon_{k}^{\mp}=3JS$ and
         $\varepsilon_{k}=0$ are superpositions of the patterns shown in panels
         (a)-(a${}^{\prime\prime}$)and (b)-(b${}^{\prime\prime}$), respectively.
         }
\label{F:covering}
\end{figure}

   We demonstrate that linear combinations of the building blocks constitute the confined states.
As an example, we focus on the rectangular regions highlighted in Figure~\ref{F:confinedreal}.
Figure~\ref{F:covering} illustrates how the Penrose lattice is covered by specific building
blocks of the confined states at $\varepsilon_{k}^{\mp}=3JS$ in the linear spin-wave
(LSW) Hamiltonian~\eqref{E:LSWdiagHam} and at $\varepsilon_{k}=0$ in the tight-binding
Hamiltonian~\eqref{E:TBdiagHam}.
The $\varepsilon_{k}^{\mp}=3JS$ confined states in the LSW Hamiltonian are covered
by type-1 to type-4 building blocks (excluding type-5 and type-6),
whereas the $\varepsilon_{k}=0$ confined states in the tight-binding Hamiltonian
are covered by all six types.

%\section[\appendixname~\thesection]{Interacting Spin-Wave Formalism with $O(S^{0})$ Quantum Corrections}
\section{Interacting Spin-Wave Formalism with $O(S^{0})$ Quantum Corrections}
\label{S:ISW}
   We express the Heisenberg Hamiltonian~\eqref{E:HeisenbergHam} in terms of
the Holstein-Primakoff transformation~\eqref{E:HPboson} and expand it in descending powers
of $S$.
The $O(S^{0})$ terms yield
\begin{align}
   &
   \mathcal{H}^{(0)}
  =-J\sum_{i\in\mathrm{A}}\sum_{j\in\mathrm{B}}l_{i,j}
   \left[
     a_{i}^{\dagger}a_{i}b_{j}^{\dagger}b_{j}
     \vphantom{\frac{1}{4}}
    +\frac{1}{4}\left(
       a_{i}^{\dagger}a_{i}a_{i}b_{j}
      +a_{i}^{\dagger}b_{j}^{\dagger}b_{j}^{\dagger}b_{j}
      +\mathrm{H.c.}
     \right)
   \right].
\label{E:H0}
\end{align}
We decompose the $O(S^{0})$ quartic Hamiltonian~\eqref{E:H0} %$\mathcal{H}^{(0)}$ 
into quadratic terms
$\mathcal{H}^{(0)}_{\mathrm{BL}}$ and normal-ordered quartic terms $:\mathcal{H}^{(0)}:$
through Wick's theorem \cite{N034714,Y094412,Y702},
\begin{align}
   &
   a_{i}^{\dagger}a_{i}b_{j}^{\dagger}b_{j}
  =
   :a_{i}^{\dagger}a_{i}b_{j}^{\dagger}b_{j}:
  +\langle b_{j}^{\dagger}b_{j} \rangle
   a_{i}^{\dagger}a_{i}
  +\langle a_{i}^{\dagger}a_{i} \rangle
   b_{j}^{\dagger}b_{j}
  +\langle a_{i}^{\dagger}b_{j}^{\dagger} \rangle
   a_{i}b_{j}
  +\langle a_{i}b_{j} \rangle
   a_{i}^{\dagger}b_{j}^{\dagger}
   \allowdisplaybreaks
   \nonumber \\
   &\qquad\qquad\quad
  -\langle a_{i}^{\dagger}a_{i} \rangle
   \langle b_{j}^{\dagger}b_{j} \rangle
  -\langle a_{i}^{\dagger}b_{j}^{\dagger} \rangle
   \langle a_{i}b_{j} \rangle,
   \allowdisplaybreaks
   \nonumber \\
   &
   a_{i}^{\dagger}a_{i}a_{i}b_{j}
  =
   :a_{i}^{\dagger}a_{i}a_{i}b_{j}:
  +2\left(
     \langle a_{i}^{\dagger}a_{i} \rangle
     a_{i}b_{j}
    +\langle a_{i}b_{j} \rangle
     a_{i}^{\dagger}a_{i}
    -\langle a_{i}^{\dagger}a_{i} \rangle
     \langle a_{i}b_{j} \rangle
   \right),
   \allowdisplaybreaks
   \nonumber \\
   &
   a_{i}^{\dagger}b_{j}^{\dagger}b_{j}^{\dagger}b_{j}
  =
   :a_{i}^{\dagger}b_{j}^{\dagger}b_{j}^{\dagger}b_{j}:
  +2\left(
     \langle a_{i}^{\dagger}b_{j}^{\dagger} \rangle
     b_{j}^{\dagger}b_{j}
    +\langle b_{j}^{\dagger}b_{j} \rangle
     a_{i}^{\dagger}b_{j}^{\dagger}
    -\langle a_{i}^{\dagger}b_{j}^{\dagger} \rangle
     \langle b_{j}^{\dagger}b_{j} \rangle
   \right),
\label{E:Wickdecomp}
\end{align}
where $\langle \cdots \rangle$ denotes a quantum average in the up-to-$O(S^{0})$ magnon vacuum.
We define the up-to-$O(S^{0})$ bilinear interacting spin-wave (ISW) Hamiltonian
\begin{align}
   \mathcal{H}^{[0]}
  \equiv
   \mathcal{H}^{(2)}
  +\mathcal{H}^{(1)}
  +\mathcal{H}_{\mathrm{BL}}^{(0)}
  =
   \mathcal{H}^{[1]}
  +\mathcal{H}_{\mathrm{BL}}^{(0)}.
\label{E:ISWHam}
\end{align}

   Let us introduce the vector representation of the Holstein-Primakoff bosons
\eqref{E:HProwvector} and matrices $\mathbf{Z}_{\mathrm{AA}}^{(0)}$,
$\mathbf{Z}_{\mathrm{BB}}^{(0)}$, $\mathbf{C}_{\mathrm{AB}}^{(0)}$ of dimensions
$L_{\mathrm{A}}\times L_{\mathrm{A}}$, $L_{\mathrm{B}}\times L_{\mathrm{B}}$, and
$L_{\mathrm{A}}\times L_{\mathrm{B}}$, respectively, as
\begin{align}
   &
   \left[\mathbf{Z}_{\mathrm{AA}}^{(0)}\right]_{i,i'}
  =-\delta_{i,i'}\sum_{j\in\mathrm{B}}l_{i,j}
   \left[ \langle b_{j}^{\dagger}b_{j} \rangle
%         +\frac{1}{2}\left( \langle a_{i}b_{j} \rangle
%                           +\langle a_{i}^{\dagger}b_{j}^{\dagger} \rangle \right)
         +\mathrm{Re}\left(\langle a_{i}b_{j} \rangle\right)
   \right],
   \allowdisplaybreaks
   \nonumber \\
   &
   \left[\mathbf{Z}_{\mathrm{BB}}^{(0)}\right]_{j,j'}
  =-\delta_{j,j'}\sum_{i\in\mathrm{A}}l_{i,j}
   \left[ \langle a_{i}^{\dagger}a_{i} \rangle
%         +\frac{1}{2}\left( \langle a_{i}b_{j} \rangle
%                           +\langle a_{i}^{\dagger}b_{j}^{\dagger} \rangle \right)
         +\mathrm{Re}\left(\langle a_{i}b_{j} \rangle\right)
   \right],
   \allowdisplaybreaks
   \nonumber \\
   &
   \left[\mathbf{C}_{\mathrm{AB}}^{(0)}\right]_{i,j}
  =-l_{i,j}
   \left[ \langle a_{i}b_{j} \rangle
         +\frac{1}{2}\left( \langle a_{i}^{\dagger}a_{i} \rangle
                           +\langle b_{j}^{\dagger}b_{j} \rangle \right)
   \right].
\label{E:H0matrixelement}
\end{align}
Then, the $O(S^{0})$ quadratic terms $\mathcal{H}_{\mathrm{BL}}^{(0)}$ can be written as
\begin{align}
   &
   \mathcal{H}_{\mathrm{BL}}^{(0)}
  =
   \varepsilon^{(0)}
  +J\left[ \bm{a}^{\dagger}, {}^{t}\bm{b} \right]
   \left[
   \begin{array}{c|c}
     \mathbf{Z}_{\mathrm{AA}}^{(0)} & \mathbf{C}_{\mathrm{AB}}^{(0)}
    \\ \hline
     \mathbf{C}_{\mathrm{AB}}^{(0)\dagger} & \mathbf{Z}_{\mathrm{BB}}^{(0)}
   \end{array}
   \right]
   \left[
   \begin{array}{c}
     \bm{a}
    \\
     {}^{t}\bm{b}^{\dagger}
   \end{array}
   \right]
  \equiv
   \varepsilon^{(0)}
  +J\bm{d}^{\dagger}\mathbf{N}\bm{d},
\label{E:H0matrix}
\end{align}
where the $O(S^{0})$ constant $\varepsilon^{(0)}$ is given by
\begin{align}
   &
   \varepsilon^{(0)}
  =J\sum_{i\in\mathrm{A}}\sum_{j\in\mathrm{B}}l_{i,j}
   \left[
     \langle a_{i}^{\dagger}a_{i} \rangle
    +\frac{1}{2}\left( \langle a_{i}b_{j} \rangle
                      +\langle a_{i}^{\dagger}b_{j}^{\dagger} \rangle \right)
    +\langle a_{i}^{\dagger}a_{i} \rangle\langle b_{j}^{\dagger}b_{j} \rangle
    +\langle a_{i}^{\dagger}b_{j}^{\dagger} \rangle\langle a_{i}b_{j} \rangle
\right.
   \allowdisplaybreaks
   \nonumber \\
   &\qquad\qquad\qquad\quad
   \left.
    +\frac{1}{2}\left( \langle a_{i}^{\dagger}a_{i} \rangle
                      +\langle b_{j}^{\dagger}b_{j} \rangle \right)
                \left( \langle a_{i}^{\dagger}b_{j}^{\dagger} \rangle
                      +\langle a_{i}b_{j} \rangle \right)
   \right].
\label{E:epsH0}
\end{align}
Thus, the bilinear ISW Hamiltonian in matrix form is given by
\begin{align}
   \mathcal{H}^{[0]}
  =\mathcal{H}^{(2)}
  +\varepsilon^{(1)}
  +\varepsilon^{(0)}
  +J\bm{d}^{\dagger}\left(S\mathbf{M}+\mathbf{N}\right)\bm{d}.
\label{E:ISWHammatrix}
\end{align}
By replacing the matrix part $JS\mathbf{M}$ of the LSW Hamiltonian~\eqref{E:LSWHammatrix}
with $J(S\mathbf{M}+\mathbf{N})$ and carrying out the Bogoliubov
transformation~\eqref{E:Bogoliubov}, the ISW Hamiltonian~\eqref{E:ISWHammatrix} can be
diagonalized in terms of the quasiparticle magnons up to $O(S^{0})$, given by
\begin{align}
   &
   \mathcal{H}^{[0]}
  =\mathcal{H}^{(2)}
  +\varepsilon^{(1)}
  +\varepsilon^{(0)}
  +\sum_{k_{+}=1}^{L_{+}}\varepsilon_{k_{+}}^{+}
  +\sum_{\sigma=\mp}\sum_{k_{\sigma}=1}^{L_{\sigma}}
   \varepsilon_{k_{\sigma}}^{\sigma}
   \alpha_{k_{\sigma}}^{\sigma\dagger}\alpha_{k_{\sigma}}^{\sigma}.
\label{E:ISWdiagHam}
\end{align}

   The expressions for the Bogoliubov transformation and the site-resolved density of states
in the ISW formalism are identical in form to Equations~\eqref{E:Bogoliubov} and \eqref{E:SWDOS}
for the LSW case, respectively.
The matrix elements of the Bogoliubov transformation, however, differ between the two cases.
This contrasts with collinear antiferromagnets on two-dimensional bipartite periodic lattices
with a single coordination number, such as the quadricoordinated square lattice and
tricoordinated honeycomb lattice, where the Bogoliubov transformation is common to both the
LSW and ISW formalisms.
The difference in the present case stems from the quasiperiodic structure of the Penrose
lattice with various coordination numbers.

\end{document}